\documentclass[fleqn,usenatbib]{mnras}

\usepackage{newtxtext,newtxmath}
\usepackage{orcidlink}

\usepackage[T1]{fontenc}
\DeclareUnicodeCharacter{2212}{-}

\DeclareRobustCommand{\VAN}[3]{#2}
\let\VANthebibliography\thebibliography
\def\thebibliography{\DeclareRobustCommand{\VAN}[3]{##3}\VANthebibliography}

\usepackage{graphicx}	
\usepackage{amsmath}	
\usepackage{multicol}        
\usepackage{gensymb}
\usepackage{color}
\usepackage[flushleft]{threeparttable}
\usepackage{xspace}
\usepackage{ulem}
\usepackage{float}
\usepackage{afterpage}

\newcommand{\bcmi}{$\beta$ CMi\xspace}

\newcommand{\aeri}{$\alpha$ Eri\xspace}
\newcommand{\beatlas}{\textsc{BeAtlas}\xspace}
\newcommand{\hdust}{\textsc{hdust}\xspace}
\newcommand{\halp}{H$\alpha$\xspace}
\newcommand{\ttms}{$t/t_{\mathrm{MS}}$\xspace}
\newcommand{\req}{$R_{\mathrm{eq}}$\xspace}
\newcommand{\vsini}{$v\sin{i}$\xspace}

\usepackage{newtxtext,newtxmath}
\usepackage{verbatim}
\usepackage{subfig}

\usepackage{soul}

\usepackage{pifont}

\title[BeAtlas: A grid of synthetic spectra for Be stars]{Bayesian sampling with BeAtlas, a grid of synthetic Be star spectra \\
I. Recovering the fundamental parameters of \aeri and \bcmi}

\author[A.~C.~Rubio et al.]
{A.~C.~Rubio$^{1, 2}$\thanks{E-mail: amanda.rubio@usp.br}\orcidlink{https://orcid.org/0000-0002-2490-1562},
A.~C.~Carciofi$^{1}$\orcidlink{https://orcid.org/0000-0002-9369-574X},
P.~Ticiani$^{1}$\orcidlink{https://orcid.org/0000-0002-4808-7796},
B.~C.~Mota$^{1}$,
R.~G.~Vieira$^{3}$,
D.~M.~Faes$^{4}$\orcidlink{https://orcid.org/0000-0001-8603-8031},
\newauthor
M. Genaro$^{5}$\orcidlink{https://orcid.org/0000-0003-3461-1929},
T.~H.~de Amorim$^{1}$\orcidlink{https://orcid.org/0000-0001-5563-6629},
R. Klement$^{6}$\orcidlink{https://orcid.org/0000-0002-4313-0169},
I. Araya$^{7}$\orcidlink{https://orcid.org/0000-0002-8717-7858},
C. Arcos$^{8}$\orcidlink{https://orcid.org/0000-0002-4825-4910},
M. Curé$^{8}$\orcidlink{https://orcid.org/0000-0002-2191-8692},  
\newauthor
A. Domiciano de Souza$^{9}$,
C. Georgy$^{10}$\orcidlink{https://orcid.org/0000-0003-2362-4089},
C. E. Jones$^{11}$\orcidlink{https://orcid.org/0000-0001-9900-1000},
M. W. Suffak$^{11}$\orcidlink{https://orcid.org/0000-0003-0696-2983},
A. C. F. Silva$^{1}$
\\
$^{1}$Instituto de Astronomia, Geof\'{\i}sica e Ci\^encias Atmosf\'ericas, Universidade de S\~ao Paulo, Rua do Mat\~ao 1226, 05508-900  S\~ao Paulo, Brazil \\
$^{2}$ European Organisation for Astronomical Research in the Southern Hemisphere (ESO), Karl-Schwarzschild-Str.\ 2, 
85748 Garching b.\ M\"unchen, Germany \\
$^{3}$ Departamento de Física, Universidade Federal de Sergipe, Av. Marechal Rondon, S/N, 49100-000, São Cristóvão, SE, Brazil\\
$^{4}$ National Radio Astronomy Observatory, 1003 Lopezville Road, Socorro, NM 87801, USA \\
$^{5}$ Instituto de Física, Universidade de São Paulo, Rua do Matão 1371, 05508-090 São Paulo, Brazil \\
$^{6}$ European Organisation for Astronomical Research in the Southern Hemisphere (ESO), Casilla 19001, Santiago 19, Chile \\
$^{7}$ Centro de Óptica e Información Cuántica, Vicerrectoría de Investigación, Universidad Mayor, Santiago, Chile \\
$^{8}$ Instituto de Física y Astronomía, Facultad de Ciencias, Universidad de Valparaíso, Casilla 5030, Valparaíso, Chile \\
$^{9}$ Université Côte d’Azur, Observatoire de la Côte d’Azur, CNRS, UMR7293 Lagrange, 28 Av. Valrose, F-06108 Nice Cedex 2, France \\
$^{10}$ Geneva Observatory, Geneva University, Chemin des Maillettes 51, 1290 Versoix, Switzerland \\
$^{11}$ Department of Physics and Astronomy, Western University, London, ON N6A 3K7, Canada \\
}

\date{Accepted XXX. Received YYY; in original form ZZZ}

\pubyear{2023}

\begin{document}
\label{firstpage}
\pagerange{\pageref{firstpage}--\pageref{lastpage}}
\maketitle

\begin{abstract}

Classical Be stars are fast rotating, near main sequence B-type stars. The rotation and the presence of circumstellar discs  profoundly modify the observables of active Be stars. Our goal is to infer stellar and disc parameters, as well as distance and interstellar extinction, using the currently most favoured physical models for these objects.
We present \beatlas, a grid of $61\, 600$ NLTE radiative transfer models for Be stars, calculated with the \hdust code. The grid was coupled with a Monte Carlo Markov chain code to sample the posterior distribution. We test our method on two well-studied Be stars, \aeri and \bcmi, using photometric, polarimetric and spectroscopic data as input to the code.
We recover literature determinations for most of the parameters of the targets, in particular the mass and age of \aeri, the disc parameters of \bcmi, and their distances and inclinations. The main discrepancy is that we estimate lower rotational rates than previous works. We confirm previously detected signs of disc truncation in \bcmi and note that its inner disc seems to have a flatter density slope than its outer disc. The correlations between the parameters are complex, further indicating that exploring the entire parameter space simultaneously is a more robust approach, statistically.
The combination of \beatlas and Bayesian-MCMC techniques proves successful, and a powerful new tool for the field: the fundamental parameters of any Be star can now be estimated in a matter of hours or days.
\end{abstract}







\section{Be stars in the era of Big Data}\label{sect:introduction}




Over the last decade, astrophysics has been flooded with all types of observational data. At the same time, high performance computing clusters grew more powerful, smaller and cheaper, calculating ever more sophisticated physical models. A single human, or even dozens, cannot go through and make sense of all the data and models available today. In order to organise, classify, analyse and systematically compare data and models, methods such as Monte Carlo, machine learning, artificial intelligence, deep learning have proven invaluable, and are now indispensable tools for astronomers {\citep[for a review, see][]{baron2019}}.

Among the objects with a profusion of observational data are {classical} B emission (Be) stars: Main Sequence (MS) stars with large rotation rates, characterised by emission lines in the Balmer hydrogen series, infrared excess and linear polarisation. These observational characteristics originate from a gaseous circumstellar disc formed from matter expelled episodically and usually unpredictably by the central star. Be stars are popular targets for many reasons. Firstly, many are bright: four of the seven sisters on the Pleiades cluster are Be stars. Achernar (\aeri), with $V = 0.46$ is the 9th brightest star in the night sky, while $\eta$ Cen ($V = 2.35$) is the 79th. In fact, the first detection of an emission line in a star was on the prototypical Be star $\gamma$ Cas \citep{secchi1866}, the 89th brightest star in the sky. Secondly, most Be stars are rather variable in photometry and spectroscopy, in time-scales from hours to decades. There are tens or so Be stars with observing histories ranging more than a century. Aside from them, large photometric surveys, such as OGLE and \textit{TESS}, are bound to have a large number of Be stars in them simply because Be stars are common: in the Galaxy, 
{they represent 17 per cent of B stars, going up to 34 per cent for spectral type B1 \citep{zorec1997}},
and in open clusters at lower metallicities, the lower fraction of Be stars can be as high as 32 per cent 
\citep{bodensteiner2020}.

%

Finding the fundamental parameters of Be stars (i.e., mass, size, rotation rate) is challenging due to the effects of their fast rotation (oblateness and gravity darkening), the disc and its variability. The star and the disc are coupled -- the apparent stellar parameters are changed by the disc, while the disc structure depends on the stellar characteristics. Some Be stars, such as Achernar \citep{domiciano2014}, have had their intrinsic properties measured with great precision via interferometry, but the fundamental parameters of the majority of Be stars are still poorly known, with estimates in the literature differing by several spectral sub-types, and sometimes not even being recognised as Be stars (e.g.,  $\nu$ Gem - B5V, OB, B7IV, B6III; $\alpha$ Col - B9Ve, B7III, B5p, B8Vn, B7IV, B8Vevar; $\beta$ Psc - B6Ve, B4V, B5p, B7IV/V, B5Ve)\footnote{\citet{buscombe}, \citet{nugem2}, \citet{nugem3}, \citet{lesh}, \citet{acol1}, \citet{arcos}, \citet{cp1993}, \citet{acol4}, \citet{acol5}, \citet{buscombe}, \citet{lesh}, \citet{arcos}, \citet{cp1993}, \citet{betaps4}, \citet{buscombe}, respectively.}. 

Efforts to model Be star discs have found success after the viscous decretion disc (VDD) model of \citet{lee1991} began to be systematically tested against observations in the last two decades or so. {Being an $\alpha$-disc} \citep{shakura1973}, a VDD builds from inside-out through the viscous shear between layers of material thrown into orbit by the central star. While the engine of the mass ejection mechanism itself is still unclear, models have been generally successful to describe the structure of the disc at large. Works such as \citet{carciofi2006b}. \citet{tycner2008} and \citet{klement2015}
created realistic physical models for Be stars in the VDD framework using the non-local thermodynamic equilibrium (non-LTE) radiative transfer codes \textsc{bedisk} \citep{sigut2007} and \hdust \citep{carciofi2006} and successfully compare the synthetic observables with data. 

The solid foundation of the VDD model opens the door to the systematic modelling of Be stars, using their wealth of data to obtain more reliable parameters of individual stars or populations of stars. 
However, detailed modelling of even one star is a time-consuming effort. The usual approach is to compare models to observational data by manual searches through the parameter space. To date, the most comprehensive model of a Be star is \citet{klement2015, klement2017}'s work with \bcmi. The authors use a tailor-made grid of dozens of \hdust radiative transfer models and a wide range of observations, such as photometric, polarimetric, interferometric and spectroscopic data. This approach, however, misses nuances of the coupled star+disc system, as many parameters are fixed and their degeneracies are, {in many cases}, not accounted for. The results of ``manual'' searches cannot be blindingly trusted as they overlook the correlations between the physical parameters of the system, and may entirely miss regions of the parameter space that provide adequate (or even better) descriptions of the observations. {Furthermore, calculating radiative transfer models in non-local thermodynamic equilibrium (NLTE), such as is done by \hdust, is a slow and computationally expensive process.}


Even with so much data available, it is difficult to study large samples of Be stars because of their variability and diversity. The favoured approach so far has been to use simplified models rather than detailed radiative transfer calculations, as is the case of \citet{vieira2015,vieira2017}. Although 
this approach may provide important insights on the properties of the sample, they also suffer from biases originating from the simplifications to the model.

Data are available; a well-tested, substantiated physical model exists. To connect them and contemplate all the drawbacks of traditional modelling, we need 1) an extensive grid of detailed, state-of-the-art radiative transfer models that consider all quirks of Be stars and is cemented in the VDD theory and 2) a statistical engine to compare observational data and models, considering the coupled nature of the system. In this work, we provide and test such an alternative modelling method. We present \beatlas, a comprehensive grid of radiative transfer Be star models that covers all the relevant stellar and disc parameters present in the Be class. We compare models to observational data using Bayesian Monte Carlo Markov chain (MCMC) methods, also taking prior information on the Be star into account, such as parallax measurements.
{The analysis is done simultaneously for all parameters, so that the degeneracy of the {model parameters} is fully considered, and their correlations can be quantified. The results consist not in a set of best-fitting parameters, but in  probability density functions (PDFs) for each parameter, allowing for robust error estimates. We apply this method to two well-studied Be stars, \aeri and \bcmi, and compare our results with their rich literature.}

\section{Overview of the VDD model}
\label{sect:vdd_model}



The observational characteristics of Be discs are well established: continuum excess 
(starting in the visible and growing towards the infrared and radio domain), linear polarisation and double peaked emission lines (when the disc is seen at inclinations different than pole-on). 
Over the years, observations made it clear that the disc must be relatively thin, 
and rotating in a quasi-Keplerian fashion (\citealt{porter1997}, \citealt{okazaki1991}; more recently, \citealt{meilland2007}, \citealt{oudmaijer2010}).
In the 90's, several disc models were proposed to explain Be discs, such as the wind compressed disc (WCD) \citep{bjorkman1993} and the VDD model (\citealt{lee1991}; see also \citealt{porter1999}; \citealt{okazaki2001}; \citealt{bjorkman2005}; \citealt{krticka2011}). It soon became clear that only the VDD model could provide an adequate description of the global picture presented by observations, and it has been the favoured model for Be star discs ever since. 
The VDD has been successfully employed in the study of individual Be stars, 
modelling either snapshots \citep[e.g.,][]{carciofi2006b, tycner2008, klement2015} or the temporal evolution of the disc \citep[e.g.,][]{carciofi2012,ghoreyshi2018,ghoreyshi2021}.
Studies of samples of Be stars were also carried out \citep[e.g.,][]
{Touhami2013, vieira2015, vieira2017,rimulo2018}.



The disc is formed because Be stars expel matter into orbit. 
The mechanism behind this mass ejection is not fully defined, although fast rotation, ubiquitous in Be stars, plays an {important} part. However, since Be stars are fast, but not critical rotators \citep[see][for a review]{rivinius2013}, the mass ejection needs additional elements to be {triggered and} sustained; common suggestions are non-radial pulsations \citep{baade2016} and small-scale magnetic fields \citep{cassinelli2002, brown2008}. 



In the VDD model, once the material is ejected and enters orbit, turbulent viscosity transports mass and angular momentum (AM) outwards, building up the disc. A VDD is similar to protostellar and accretion discs \citep{pringle1981}, the difference being that VDDs switch between \textit{outflowing} and \textit{inflowing}, while pre-MS and accretion discs are always \textit{inflowing}. These discs are also completely dust- and molecule-free, being composed mostly of hydrogen.

If one assumes that mass is lost from the {stellar equator {uniformly and} at a constant rate, and that the thin disc approximation is valid \citep[][for further details]{bjorkman2005}, 
the viscous diffusion effectively reduces to a one-dimensional problem, with the density and velocity depending only on the radial distance to the star
\footnote{{The thin disc approximation valid at near and moderate distances from the star, where we are focused. As the disc flares, a two-dimensional density profile becomes important far away from the central star (i.e., hundreds of stellar radii). For more details, refer to \citet{kurfurst2018}, specially their Eq.~37 and its derivation.}}
{If we further assume that the disc is radially and vertically isothermal, and that mass is lost by the central star at a constant rate for a sufficiently long time }\citep[see][]{haubois2012}, an analytical solution for the disc structure can be {derived}. Considering this, \citet{bjorkman2005} obtained the following expression for the surface density



\begin{equation}\label{eq:VDDsigma}
    \Sigma(r) = \frac{\dot{M} v_{\rm orb} R_{\rm eq}^{1/2}}{3 \pi \alpha \, c_s^2 \,r^{3/2}} \left[\left(\frac{R_0}{r}\right)^{1/2} - 1 \right]\,,
\end{equation}
where $R_{\rm eq}$ is the equatorial radius of the central star, $v_{\rm orb}$ is the Keplerian orbital velocity at the equator, $\dot{M}$ is the rate of mass flowing out of the central star, $\alpha$ is the Shakura-Sunyaev viscosity parameter, $R_0$ is an integration constant, associated with the outer, torque-free boundary condition, and the sound speed in the disc is 
{
$c_s = ({kT/m_\mathrm{H} \mu})^{1/2}$} ($T$ is the disc temperature, $k$ is the Boltzmann constant, $m_\mathrm{H}$ is the hydrogen mass, and $\mu$ is the mean molecular weight of the gas). Given that $R_0 \gg R_{\rm eq}$, Eq.~\ref{eq:VDDsigma} assumes a simple functional form $\Sigma(r) \propto \Sigma_0 \, r^{-2}$, where $\Sigma_0$ is the surface density at the base of the disc. 

In the absence of radiative forces and nearby stellar companions, {and assuming that the material is in vertical hydrostatic equilibrium}, the vertical (i.e., along the direction orthogonal to the disc plane) density distribution is a simple Gaussian with a scale height
controlled solely by gas pressure and by the gravitational field of the Be star: as the disc grows further from the star, it will flare. Thus, the scale height $H$ has the form 
\begin{equation}
    H = H_0 \left( \frac{r}{R_{\rm eq}} \right)^\beta, 
\label{eq:scaleheight}
\end{equation}
where $H_0 = c_s R_{\rm eq}/v_{\rm orb}$ and $\beta$, the flaring exponent, is 1.5 \citep{bjorkman2005}. 
Finally, we can obtain the equation for the volume mass density, $\rho$ of an isothermal VDD by considering that the surface density $\Sigma$ is the vertical integral of the density. This allow us to write $\rho$ in terms of the scale height and the surface density
\begin{equation}
\rho (r,z) = \frac{\Sigma_0}{H \sqrt{2\pi}} \left(\frac{r}{R_{\rm eq}}\right)^{-2}\exp\left[-\frac{z^2}{2\,H^2}\right] \propto \rho_0 \,  r^{-n},
\label{eq:rhovdd}
\end{equation}
with $n = 3.5$ and $\rho_0$ is the base volume density of the disc. 
In this work we present our results in terms of the number density $n_0$, which relates to the volume density as $n_0 = \rho_0 \, N_A/\mu$, where $N_A$ is Avogadro's number.


\subsection{Deviations from a power-law density}\label{subsec:deviations}

The above considerations and the power law approximation of Eq. \ref{eq:rhovdd} are valid for an isothermal, steady-state VDD. Real Be discs, however, are rarely either. Be discs are famously variable in many time-scales. They are  dependent on injection of mass and AM on their base to exist, and variations on the mass injection rate affect their properties in complex ways. At its most dramatic, if mass injection stops for a long period (larger than the typical viscous timescales), the disc completely dissipates and the Be star is seen as a regular (but fast rotating) B star. The inner disc dissipates faster, falling back into the central star as it is no longer able to support itself without the injection of AM. The outer disc still receives AM from the decaying inner disc, and can therefore survive for longer, but dissipates {with time}. These disc build-up and dissipation events happen in the viscous diffusion time-scale, given by 
\begin{equation}
    \tau_{\nu} = \frac{v_{\rm orb}}{\alpha c_s^2} \sqrt{R_{\rm eq} \, r} \sim 20 \, \mathrm{years}\, \frac{0.01}{\alpha} \sqrt{\frac{r}{R_{\rm eq}}}\,,
\end{equation}
meaning that Be discs can be completely built and lost in timescales of months to years given that estimates of the viscous parameter $\alpha$ put it in the ballpark of a few tenths \citep{rimulo2018,ghoreyshi2018}. 

A Be star that is in the process of either building or losing its disc will not have a volume density that follows the $n = 3.5$ prescription. Given their inside-out growth and dissipation, discs that are building-up have
$n$ higher than 3.5, as the inner part of the disc fills up faster than the outer part. In dissipation, we have the opposite ($n < 3.5$), as the inner disc is also quicker to be depleted than the outer disc \citep[][their Figs. 1 and 3]{haubois2012}. 
In fact, the situation is even more complex than depicted above, because the models of \citet{haubois2012} show that $n$ actually varies strongly with the  distance to the star (e.g., their Figs.~3 and 6).
\citet{vieira2017} used this to connect their estimates of $n$ for their sample of 80 Be stars to the current state of the disc, with $3.0 < n < 3.5$ being considered a region of disc stability. 
Most of their sample had $n<3.5$, suggesting these discs were dissipating at the time of the observations, whereas only 45 per cent of the discs were either in build-up (24 per cent) or stable (21 per cent).

Another factor that can cause $n$ to deviate from 3.5 is the presence of a binary companion. 
As massive stars, early-type\footnote{Spectral classes B0-B3} Be stars are inserted in the larger context of massive star multiplicity: around $80$ per cent have interactions with a companion in some point of their lives \citep{Sana2012}. Multiplicity fraction decreases with mass \citep{janson2012, sota2011}, but late-type\footnote{Spectral classes B4-B9} Be stars have been found to have companions as well \citep[e.g., $\kappa$ Dra --][]{klement2022}. \citet{oudmaijer2010}, {for instance}, finds a binary fraction of $30 \pm 8$ per cent for Be stars specifically {(their method could probe separations from 0.1 to 8 arcsec, and magnitude differences up to 10 mag)}.
The binary fraction among Be stars is not well constrained, with suggestions ranging from {$5$} \citep{vanbever1997} to nearly $100$ per cent \citep{shao2014, klement2019}. 
\citet{okazaki2002}, in their smoothed particle hydrodynamics simulations of binary Be stars, find that when the companion has a close enough orbit, it tidally affects the disc, which can lead to disc truncation and to the accumulation of matter inside of its orbit, making the {radial density falloff} much shallower ($n < 3.5$ -- \citealt{panoglou2016}, \citealt{cyr2017}). This accumulation effect is stronger for smaller binary separations, smaller viscosities and larger mass ratios.


The radial density exponent can also deviate from the isothermal $n = 3.5$ if the disc has a temperature structure that is highly \textit{non}-isothermal {\citep{carciofi2008, mcgill2013, kurfurst2018}}. As the sound speed in the disc (and hence the viscosity) {depends} on the temperature, the entire density structure of the disc can change profoundly if $T$ varies either in time or space.
For instance, \citet{carciofi2008} find that in a non-isothermal disk, albeit steady-state, $n$ can range between 2 and 6 (their Fig.~4).
The disc structure also depends on the viscosity parameter $\alpha$, which may also vary (independently of the temperature) throughout the disc and also in time \citep{ghoreyshi2021}. {In fact, it is not straightforward to distinguish the effects that a variable $\alpha$ or a variable $T$ would have on the disc. Thus, it is more useful to think of their combined effect, i.e., how the disc responds to variations of $\alpha T$.} 

{{Summarising}, $n$ can vary in time according to the dynamical status of the disc and the tidal influence of a companion, and vary radially due to non-isothermality and variations in viscosity. Therefore, the power law density of Eq.~\ref{eq:rhovdd} is an approximation for the staggering majority of Be stars.}
At the same time, it is a 
useful approximation as it allows measuring -- via detailed modelling -- the (average) disk density slope at a given moment in time, irrespectively of how this slope was realised.

\section{The Central Star}\label{sect:central_star}




Rotation is an important ingredient in the formation, structure and evolution of stars in the MS {\citep[e.g.,][] {maeder2009}}. 
\citealt{ekstrom2012} shows that the evolutionary tracks for rotating and non-rotating hot stars differ significantly, with rotators having chemically enhanced stellar surfaces and enjoying longer MS lifetimes\footnote{This is only false when a star is very metal poor, and the dominant effect of rotation is the helium diffusion to the outer envelope, making the star overluminous and thus reducing lifetime \citep{meynet2002}.} due to rotational mixing bringing more hydrogen to the core {\citep{georgy2013}}. These effects are especially relevant for Be stars since they are the fastest rotators in the MS \citep{Cranmer2005}. 

{
The causes of the spin up of Be stars are still not clear. Three main channels have been proposed for how a B star could acquire this fast rotation. 
The first is that they are simply born as very fast rotators, and remain so for the rest of their MS lifetime \citep{zorec1997}. The second idea is that the spin up of the star is a consequence of the stellar evolution along the MS, with the acceleration of the outer layers of the star resulting from contraction of its core \citep{ekstrom2008,granada2013}. The third option is binary evolution. \citet{pols1991} considered the possibility that Be stars are formed through the evolution of mass-transfer in close binaries. In this scenario there are two intermediate mass interacting stars, with one of them donating mass and AM to its companion. This companion, formerly a regular B star, becomes a Be star, while the donor evolves into a He star (or sub-dwarf O) or a compact object (WD, NS or BH).} {If Be stars are indeed the product of post-mass transfer systems, their evolution was surely influenced by the companion. However, including evolutionary models of binaries (which carry many parameters and uncertainties on their physical bases) is outside of the scope of this work; we consider single star evolution models for Be stars. }


The Roche equipotential formalism is frequently used to describe the rotating stellar geometry \citep[][and references therein]{cranmer1996}. The star is deformed according to the rotation rate,

\begin{equation}
    W = \frac{v_\mathrm{rot}}{v_\mathrm{orb}},
    \label{eq:rotation_rate}
\end{equation}

\noindent where $v_\mathrm{rot}$ is the rotational velocity at the equator \citep{rivinius2013}. For critically rotating stars, $W = 1$. Stars with high values of $W$ ($> 0.5$), such as Be stars, are subject to rotational deformation \citep[e.g.,][]{1966ApJ...146..152C,maeder2009}, enhanced chemical mixing, and gravity darkening \citep{1963ApJ...138.1134C, vonzeipel1924}. A Be star is therefore flattened along its rotational axis, with hotter (and thus brighter) poles and a dimmed equator. 
This has a direct consequence on the thermal structure of the disc
\citep[e.g.,][]{2011ApJ...743..111M}, influencing the observables {\citep{townsend2004, fremat2005}}. Therefore, any accurate Be star model must take gravity darkening and oblateness into account.

\section{Description of Model Grid}
\label{sect:beatlas}

The \beatlas project arises in the context of grid-based modelling, routinely applied to infer parameters of physical models (e. g., \citealt{choi2016}). \beatlas comprises two systematic grids of \hdust models. The \textit{photospheric grid}  (Sect.~\ref{sect:photospheric_grid}) covers the typical range of stellar parameters (Sects.~\ref{sect:central_star} and \ref{subsect:modeling_central_star}) from spectral types of early A to late {O}, from non-rotating ($W=0$) to near-critical rotation. The \textit{disc grid} (Sect.~\ref{sect:disk_grid}) contains models of stars surrounded by discs, computed according to the simplified VDD formulation of Sect. \ref{sect:vdd_model} and covering the typical range of stellar and disc parameters associated with the Be phenomenon. 
Details on the synthetic observables included in both grids are given in Sect.~\ref{sect:grid_observables}.

\subsection{Radiative Transfer Calculations}\label{sect:hdust}


The models were calculated with the Monte Carlo radiative transfer code \hdust \citep{carciofi2004,carciofi2006, carciofi2008}. This code has been successfully applied in several previous studies to interpret multi-technique observations. Some recent examples are the interferometric study of Achernar \citep{2017A&A...601A.118D}, the computation of line profiles of binary Be stars \citep{panoglou2018}, and the interpretation of optical light curves using the VDD model \citep{rimulo2018, ghoreyshi2018}.

The code accepts arbitrary 3D geometry (i.e., density distribution) and kinematics. The circumstellar chemistry considers atomic hydrogen in full non-LTE and/or dust grains. 
The implemented gas opacity sources are Thomson scattering, bound-bound, bound-free and free-free absorptions. The code describes the star realistically, including limb darkening, gravity darkening, and rotational flattening according to the Roche model (see details in Sect.~\ref{sect:central_star}). Its outputs include the synthetic observables such as the polarised emergent spectrum (both continuum and {Hydrogen lines in emission and absorption}) and synthetic images.


\hdust is based on the Monte Carlo method, {that discretises the radiation field in a number of photon packets, here understood as a cohesive ensemble of photons that are launched by the star and interacts with the disc (i.e., is absorbed or scattered). The accuracy of the simulation} depends on the number of photon packets used \citep[e.g.,][]{wood1997}. 





\subsubsection{Modelling the central star}\label{subsect:modeling_central_star}





The \hdust code requires the following information about the central star as input parameters: mass ($M$), polar radius ($R_{\rm p}$), luminosity ($L$), rotation rate ($W$), and a gravity darkening parameter (see below). In addition, the emergent spectrum at each position of the photosphere must be specified. For that, we use the model atmospheres from \citet{kurucz1994}, while the limb darkening follows the prescription given by \citet{claret2000}.

The gravity darkening is described by the modified von Zeipel law, in the form $T_{\rm eff} \propto g^{\beta_\mathrm{GD}}$. We adopt the model presented by \citet{espinosa2011}, where the gravity darkening exponent $\beta_\mathrm{GD}$ is a function of the rotation rate. 
In this prescription, the effective temperature $T_{\mathrm{eff}}$ on the surface of the star becomes a function of $W$ and the polar angle {(colatitude)} $\theta$\footnote{The polar angle is the angle between a given point in the stellar surface and its spin axis.}, as in Eq.~31 of \citet{espinosa2011}. For a given rotation rate, a model correction factor is found using an iterative numerical procedure, and $\beta_\mathrm{GD}$ is then estimated by fitting the $\theta$ dependent profiles of $T_{\mathrm{eff}}$ and $g$ using a power law \citep[see][for more details, specially Eq.~3.17]{espinosa2014}. This theoretical approach has been validated in works such as \citet{domiciano2014}, where the predicted values of the model were compared to interferometric measurements of six rapidly rotating stars. 
By adopting this prescription for the gravity darkening, the number of free parameters for the central star reduces from five to four.





In order to connect the parameters used by \hdust to the stellar physical parameters (age, polar radius and luminosity) in a consistent way, we use the rotating stellar models of \citet{georgy2013} and \citet{granada2013}, hereafter referred to as the Geneva grid. We selected models with solar metallicity ($Z=0.014$), according to \citet{1998SSRv...85..161G}. These models cover the mass range from 1.7 to 20\,$\mathrm{M_\odot}$ and initial rotation rates from $W=0$ to $W=0.998$. Specifically for this work, $20\,\mathrm{M_\odot}$ models were computed to properly cover the high-mass end of the Be phenomenon, {following the exact same physical description that the ones in \citet{georgy2013}}.
The chosen mass range includes all typical masses of known Be stars as well as the transition to late type Oe stars \citep{negueruela2004} and to early-A stars. By using the Geneva grid, the number of free parameters for a star reduces from four to three {(mass, rotation rate and age)}.



\subsection{\beatlas: Photospheric Grid}
\label{sect:photospheric_grid}



The photospheric grid is applicable for discless (i.e., inactive) Be stars, normal B stars and Bn stars, and contains $660$  photospheric models.
(11 masses $\times$ 10 rotation rates $\times$ 6 ages). 
Each model was computed for 10 inclination angles, so the grid effectively contains 6600 emergent spectra.
The grid is summarised in Tab.~\ref{tab:stellar_beatlas_params}.
The stellar mass was chosen in the range 1.7 to 20\,$\mathrm{M_\odot}$, typically covering the spectral range from A7 to O8.5. The association between mass and spectral type are based on \citet{2005A&A...436.1049M}, {\citet{schmidt-kaler1982}}, and \citet{2004IAUS..224....1A}, for O, B and A spectral types, respectively.
The rotation rate covers the full range between a non-rotating star ($W=0$) to a nearly-critically rotating one $W=0.99$.
The chosen steps follow
$W_i = 0.99\left(i/N\right)^{1/2}$ with $i=0,\ldots ,N$ and $N=10$. As mentioned in Sect.~\ref{subsect:modeling_central_star}, there is one specific $\beta_\mathrm{GD}$ for each rotation rate, with values ranging from $\beta_{\mathrm{GD}}=0.25$ for $W=0$, and $\beta_{\mathrm{GD}}\approx 0.13$ for $W=0.99$.



In our models, the age of a star is given in terms of its MS lifetime. 
This lifetime is expressed in the Geneva grid as the amount of H present in the core of the star. 
In the MS, the age parameter ranges from \ttms$=0$ (H fraction $X \approx 72$ per cent and He fraction $Y \approx 26$ per cent) to \ttms$=1.0$ ($X \approx 0$ per cent), where $t/t_\mathrm{MS}$ represents the fraction of time already spent by the star relative to the time duration from zero-age MS to terminal-age MS. The values for \ttms shown in Tab.~\ref{tab:stellar_beatlas_params} were chosen to better sample the trajectory of the central star along its MS track, as displayed in Fig.~\ref{fig:HRphot}. From $0 \le$ \ttms $\le 1$, the conversion between \ttms to real age is mass and W dependent.
{Given a mass, a rotation rate and \ttms, the Geneva grid is interpolated 
to calculate the corresponding $R_{\rm p}$, $L$ and age.}
This grid has known numerical issues for the models of evolved stars (\ttms$ > 1$), in particular for fast rotators. 
To sidestep this problem, we included another value for the age parameter, \ttms$ = 1.25$, as an extrapolation of the MS models rather than using the problematic post-MS models. Therefore, this value of \ttms does not correspond to a real age, but is used to explore parameter ranges (size and luminosity) that are consistent with a star just leaving the MS {beyond the ones predicted by the Geneva grid for the MS.} The associated values of $R_\mathrm{p}$ and $L$ are extrapolated using an exponential function. In this manner, we can accommodate {possibilities that are not predicted, and not contemplated, by the Geneva MS grid, such as slightly evolved stars.}

\begin{figure}
    \centering
    \includegraphics[scale=0.33]{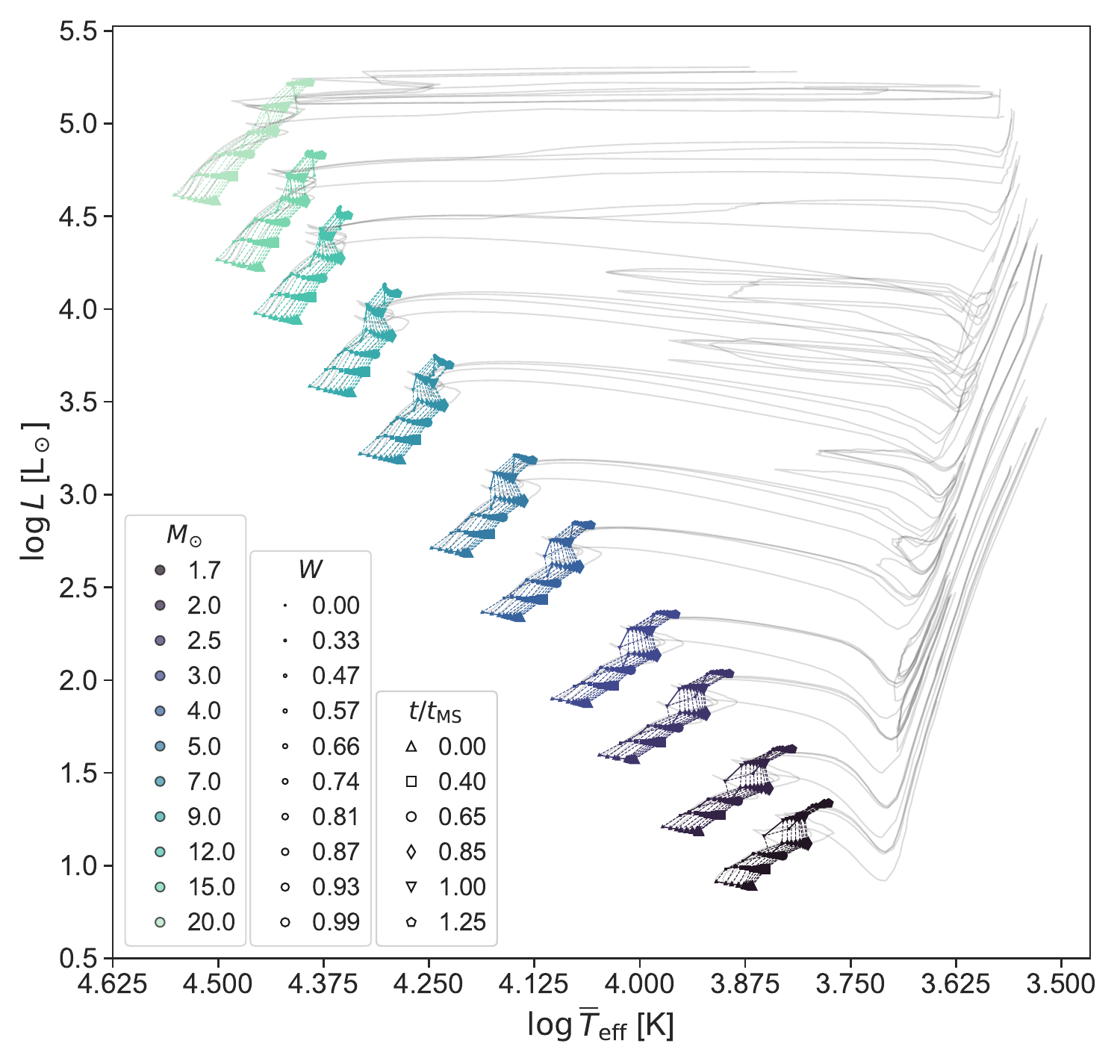}
    \caption{Hertzsprung–Russell diagram comprising the photospheric grid parameters. Masses are indicated by different marker colours. Marker sizes represent different rotation rates $W$. In grey solid lines, four complete Geneva tracks are showed, corresponding to the initial velocities of $W_\mathrm{Geneva}=0, \, 0.769, \, 0.925$ and $0.998$ for each stellar mass. Values of \ttms for each ($M$, $W$) pair are connected with dashed lines, following the tracks, indicated by different marker types; the path from ZAMS to TAMS goes from bottom left up to top right.}
    
    
    
    \label{fig:HRphot}
\end{figure}


Finally, the last parameter of the photospheric grid is the inclination angle of the stellar spin axis {with respect to the line of sight}, $i$. 
For $i$, we chose 10 values, equally spaced in $\cos i$.
As shown in the example colour-magnitude diagrams of App.~\ref{sect:observationalHR}, this parameter has an important effect on the spectral energy distribution (SED) as a whole and naturally also on the line profiles. 



\begin{table*}
\begin{threeparttable}
\centering
\small
\caption{\beatlas stellar parameters of the photospheric grid. For each rotation rate, there is only one associated value of the gravity darkening exponent. %
The non-rotating models are assigned with $\beta_{\mathrm{GD}}=0.25$, and for the critically rotating models $\beta_{\mathrm{GD}}\approx 0.13$.} 
\label{tab:stellar_beatlas_params}
\begin{tabular}{l l l l}
\hline\hline
                   &  Parameter                        & Range      &  Values                                              \\ 
            Fundamental parameters  & & & \\ \hline

$M$          &  Mass ($\mathrm{M_\odot}$)         & 1.7 - 20    &  1.7, 2, 2.5, 3, 4, 5, 7, 9, 12, 15, 20              \\
$W$                &  Rotation Rate     & 0 - 0.99 &  0.00, 0.33, 0.47, 0.57, 0.66, 0.74, 0.81, 0.87, 0.93, 0.99 \\
$t/t_\mathrm{MS}$  &  Stellar Age                       & 0 - 1.25    &  0, 0.40, 0.65, 0.85, 1, 1.25                      \\
$\cos i$           &  Inclination angle & 0 - 1  &  0, 0.11, 0.22, 0.33, 0.44, 0.55, 0.67, 0.78, 0.89, 1 \\ 
$i$           &  Inclination angle & 90 - 0  &  90.00, 83.68, 77.29, 70.73, 63.90, 56.63, 47.93 , 38.74 , 27.13,  0. \\ \\

                 Derived parameters & & & \\ \hline   
ST                 &  Spec. Type                        & A7 - O8.5     &  A7, A2, A0, B9.5, B7.5, B6, B3.5, B2.5, B1, B0.5, O8.5  \\    
$\beta_{\mathrm{GD}}$       &  Gravity darkening exponent        & 0.25 - 0.13 &  0.25, 0.23, 0.21, 0.20, 0.19, 0.18, 0.17, 0.16, 0.15, 0.13   \\
\hline
\end{tabular}
\end{threeparttable}
\end{table*}

\subsection{\beatlas: Disc Grid}
\label{sect:disk_grid}


For the disc grid, the stellar parameters {were selected based on the best representation of the} Be phenomenon. The mass range lower limit was set at $3\,\mathrm{M_{\odot}}$ 
and the upper limit of $20\,\mathrm{M_{\odot}}$ was unchanged. In total, we chose $11$ values of stellar masses for the disc grid. We restricted the number of $W$ values     to 5, beginning at $W = 0.50$ up to $W = 0.99$, since no slow rotating Be star is known to exist \citep[see the compilation presented in][their Fig.~9]{rivinius2013}. Also, the age parameter $t/t_\mathrm{MS}$ was confined to a total of 4 values, including an extrapolated non-physical value of $t/t_\mathrm{MS}=1.25$. The only unchanged parameter was the inclination angle. 
The main reason for the above choices is computation time, as a more complete stellar grid would make computing the disc grid prohibitively expensive.




Following the VDD formulation of Sect.~\ref{sect:vdd_model}, we assume that the radial disc density falloff is a power-law with the exponent, $n$, as a free parameter (Eq.~\ref{eq:rhovdd}).
Further parameters describing the disc are the volume number density ($n_0$) 
at the disc base ($r=R_{\rm eq}$) and the disc radius ($R_\mathrm{D}$).
The disc scale-height $H$ follows a simple power-law of the form of Eq. \ref{eq:scaleheight}. To compute $H_0$, the disc temperature was set to $T=0.60\,T_{\mathrm{pole}}$ following \citet{carciofi2006}, and the flaring exponent $\beta$ was fixed at the isothermal value of $1.5$.







{For each $M$ value, {a grid of} $n_0$ was built with 8 points equally {distant in logarithmic space} between a fixed lower limit and a mass-dependent upper limit. The latter corresponds to \hdust models convergence limit, adapted to an analytical approximation, described by:}

\begin{equation}
    \ln n_{0}^{\mathrm{upper}}(M)\,[\mathrm{cm}^{-3}] = \left\{\begin{array}{ll}
    a M^2 + b M + c, & M \leq 7\,\mathrm{M_{\odot}},\\ 
    d M + e, & M > 7\,\mathrm{M_{\odot}},\\
    \end{array} \right.
    \label{eq:n0upper}
\end{equation}
where $a = -0.152/\mathrm{M_{\odot}}^2$, $b = 2.51/\mathrm{M_{\odot}}$, $c = 22.09$, $d = 0.14/\mathrm{M_{\odot}}$, and $e=31.02$. {The works of \citet{vieira2017} and \citet{rimulo2018} revealed a tendency for massive stars to have much denser discs than their low mass siblings, giving an observational consistency with our convergence limitations.} 
The maximum $n_{0}$ value -- corresponding to $M=20\,\mathrm{M_{\odot}}$ -- was chosen to be $n_0 = 5 \times 10^{14} \, \mathrm{cm}^{-3}$, the highest reported in the literature for a B0 star \citep{carciofi2006}.
For the lower limit we adopted the value of $10^{11} \, \mathrm{cm}^{-3}$ for all stellar masses, which roughly corresponds to the detection limit of an emission feature (however weak) in the \halp line.





The disc radius ranges from $10$ to $100\,R_\mathrm{eq}$. In theory, Be discs can expand to hundreds of stellar radii. However, if the Be is a binary system, the presence of the companion can tidally affect the disc and truncate it. In our grid, the lower limit value of the disc radius corresponds to an estimate for the disc radius of the Be star $o$ Pup, which has the shortest period among all known binary Be stars \citep{koubsky2012}, and thus the smaller known disc radius. On the other hand, the upper limit, $100\,R_\mathrm{eq}$, corresponds to a disc size for which the disc emission starts to be negligible at all wavelengths considered in the simulation 
\citep[see Sect.~\ref{sect:grid_observables}; and also][]{vieira2015}. 


The $n$ exponent (Eq. \ref{eq:rhovdd}) range was chosen, according to \citet{vieira2017}, from 1.5 to 4.5. This range is in broad agreement with other studies from the literature \citep[e.g.,][]{1989A&A...213L..19W,Silaj2014,Touhami2014}. Recall that for an isothermal, steady-state disc, $n = 3.5$. {An overview of the parameters of the disc grid is given in Tab. \ref{tab:disk_beatlas_params}.}


\begin{table*}
\begin{threeparttable}
\centering
\small\caption{\textsc{BeAtlas} parameters of the disk grid. The values for $i$ are the same as in Tab.~\ref{tab:stellar_beatlas_params}.}


\label{tab:disk_beatlas_params}
\begin{tabular}{l l l l}
\hline\hline
                   &  Parameters                                          & Range      & Values                   \\
            Fundamental parameters  & & & \\ \hline

$M$          &  Mass ($\mathrm{M_\odot}$)                   & 3 - 20      & 3, 3.5, 4, 4.5, 5, 6, 7, 9, 12, 15, 20     \\
$W$                &  Rotation Rate                               & 0.50 - 0.99  & 0.50, 0.75, 0.85, 0.92, 0.99   \\
$t/t_\mathrm{MS}$  &  Stellar Age                                 & 0 - 1.25       & 0, 0.65, 1, 1.25                \\
$\log_{10}n_0$ (lower)        &  Disc base number density (cm$^{-3}$ - dex) & Fixed          & $11.0$\\
$\log_{10}n_0$ (upper)         &  Disc base number density (cm$^{-3}$ - dex) & $M_\odot$-dependent & $12.3$, $12.6$, $12.9$, $13.2$, $13.4$, $13.9$, $14.0$, $14.2$, $14.4$, $14.7$       \\
$R_\mathrm{D}$     &  Disc radius ($R_\mathrm{eq}$)                       & 10 - 100    & 10, 20, 40, 70, 100                            \\
$n$                &  Number density radial exponent                        & 1.5 - 4.5   & 1.5, 2.0, 2.5, 3.0, 3.5, 4.0, 4.5 \\ \\
Derived parameters & & & \\ \hline 
ST                 &  Spec. Type                                  & B9.5 - O8.5     & B9.5, B9, B7.5, B6.5, B6, B4.5, B3.5, B2.5, B1, B0.5, O8.5    \\
$\beta_{\mathrm{GD}}$       &  Gravity darkening exponent                  & 0.21 - 0.13  & 0.21, 0.18, 0.16, 0.15, 0.13   \\
\hline
\end{tabular}
\end{threeparttable}
\end{table*}





The disc grid comprises a total of $61\,600$ models (11 masses $\times$ 5 rotation rates $\times$ 4 ages $\times$ 7 density slopes $\times$ 5 disc sizes $\times$ 8 values for $n_0$), each computed for 10 inclination angles, resulting in $616\,000$ emergent spectra.
Given that each complete model can take on average 15\,h to be calculated using 64 computing cores, to complete \beatlas  approximately 15 $\times$  $61\,600$ $\times$ 64 = 59 million core-h\footnote{The core-h is a standard unit for measuring computing time in parallel computers, and is equivalent of one hour of computing time in one computing core.} are required, a challenging number even for modern computing standards.
We emphasise that the complete disc grid could not be completed for this work. Parts of the disc grid were computed, aiming at bracketing the parameter space needed to study our targets, \aeri and \bcmi (Sect.~\ref{sect:aeri} and \ref{sect:bcmi}). At the moment, {the disc grid is 32 per cent complete. }

\subsection{Observables}\label{sect:grid_observables}

To finish the specification of \beatlas, we provide a list of observables (e.g., continuum bands, spectral lines, images, etc.). 
The most efficient way to run \hdust is to compute individual runs for each spectral range of interest. To make optimal usage of this feature, we specified 22 spectral bands (first part of Tab.~\ref{tab:beatlas_observables}) covering all the way from the UV up to the radio. Each band is defined by a minimum and maximum wavelength, the number of spectral bins (i.e., its spectral resolution), their spacing (linear or logarithm), and the number of photon packets. 


We are also interested in spectral lines (both in emission and in absorption) in the optical and IR domains. These lines were selected among the most commonly used in the literature (second part of Tab. \ref{tab:beatlas_observables}).
For some observables we also compute images across pre-defined spectral channels aiming at reproducing interferometric observables. Our definitions include both decommissioned interferometers (e.g., AMBER\footnote{\url{www.eso.org/sci/facilities/paranal/instruments/amber/overview.html}} and MIDI\footnote{\url{www.eso.org/sci/facilities/paranal/decommissioned/midi.html}}), for which large number of archival observations are available, and current interferometers at ESO (GRAVITY\footnote{\url{www.eso.org/sci/facilities/paranal/instruments/gravity.html}} and MATISSE\footnote{\url{www.eso.org/sci/facilities/develop/instruments/matisse.html}}) and Georgia State Unversity's CHARA\footnote{\url{www.chara.gsu.edu/}}.







\section{MCMC Sampling}\label{sect:methodology}





A grid of models, no matter how much science has been put to it, is useless without a statistical engine {that explores the parameter space} (e.g., \citealp{klement2017}; \citealp{rimulo2018}; \citealp{bouchaud2020}; \citealp{bowman2020}). With this intent, we combine \beatlas with Bayesian statistics and Monte Carlo Markov Chain (MCMC) methods. The use of Bayesian-MCMC data analysis in astronomy has grown exponentially in recent years due to advances in both computing power and computer science, revolutionising the way we see data in science \citep[e.g.,][and references therein]{sharma2017}. 

In Bayesian formalism the posterior distribution is defined as 
\begin{equation}
p(\Theta,\alpha|D) \propto p(\Theta,\alpha) p(D|\Theta,\alpha),
\label{eq:bayesian}
\end{equation}
\noindent where $\Theta$ is a unique set of parameters that describe a model, $D$ is the data and $\alpha$ is the previously known information of the problem. As such, $p(\Theta,\alpha)$ is the \textit{a priori} distribution and $p(D|\Theta,\alpha)$ is the likelihood. 

Our goal is to find probability density functions (PDFs) of the parameters 
that can together describe a Be star given a suitable set of models, observational data 
and prior information.
To do {this}, we need a code to search through the parameter space, iteratively comparing models to data [$p(D|\Theta,\alpha)$], thus sampling the posterior distribution [$p(\Theta,\alpha|D)$]. We use a \textsc{Python} implementation of \citet{goodman2010}'s affine-invariant ensemble sampler, the high-performance open source \textsc{emcee}\footnote{Available online under the MIT License: \url{https://github.com/dfm/emcee}} \citep{foreman2013}. It is an ensemble sampler that follows a variation of the classic Metropolis-Hastings \citep{metropolis1953, hastings1970} algorithm called stretch move. Being an ensemble sampler, \textsc{emcee} uses many ``walkers'', who thread along different (but connected) chains to sample the posterior distribution. Every time one of these walkers takes a step, the other walkers are informed. The ensemble thus performs the stretch move together, reaching convergence faster (i.e., in fewer steps along the chain) than a single, lone walker could. {In every step and for every walker, a new set of parameters is proposed by \textsc{emcee}. The \beatlas synthetic observables are  linearly interpolated to this new set of parameters using the \textsc{python} \textsc{scipy}\footnote{For more details on the \textsc{scipy} package, refer to \citet{scipy}.} function \textsc{griddata} to find the synthetic observables that correspond to these parameters. The synthetic observables are then compared to the data of the target using a likelihood function.}



Although the stretch move algorithm will converge for different likelihood functions, the most commonly used one is the $\chi^{2}$ distribution. The type of likelihood depends on the observational data given to the code. For the SED from the visible to radio ($0.39 \, \mu$m, forward), we adopt a logarithmic {$\chi^2$} in the form {(\citealt{robitaille2007}, Eq.~6.; \citealt{klement2017}, Eq.~4)}

\begin{equation}
    \chi^2_{\rm mod} = \sum^{N}_{\mathrm{i} = 1} \left[ \frac{\log(F_{\mathrm{obs, i}}/F_{\mathrm{mod, i}})}{\sigma_{F_{\mathrm{obs, i}}} / F_{\mathrm{obs, i}}} \right]^2,
\end{equation}

\noindent where $F_{\mathrm{obs, i}}$ are the observed fluxes, {$\sigma_{F_{\mathrm{obs, i}}}$} are their errors, and $F_{\mathrm{mod,i}} = F_{\mathrm{mod, i}}(\Theta)$ represents the \beatlas grid of models described in Sect.~\ref{sect:beatlas} {for each wavelength $\mathrm{i}$}. The likelihood is 

\begin{equation}
\log p(D|\Theta,\alpha) \propto -0.5 \, \chi^2_{\rm mod} .
\label{eq:likelihood}
\end{equation}

The use of a logarithm $\chi^2$ offers better performance when the data has a large dynamic range: the SED of a Be star from the visible to the radio can vary 6 orders of magnitude in wavelength and 17 orders of magnitude in flux (e.g., the inset plot of Fig. \ref{fig:bcmi_full}). 
For the UV part of the SED we adopt a simple, non-logarithmic $\chi^2$, because of the smaller wavelength (and dynamical) range, in the form 

\begin{equation}
    \chi^2_{\rm mod} = \sum^{N}_{\mathrm{i} = 1} \left[ \frac{F_{\mathrm{obs, i}}- F_{\mathrm{mod, i}}}{\sigma_{F_{\mathrm{obs, i}}}} \right]^2 .
\end{equation}

Thus, if we give the code the complete SED, from UV to radio, as input data, then the two likelihoods are combined by summing the reduced $\chi^2$ of each, and multiplying by the total number of data points in the complete SED. When an \halp profile or polarisation data are the input, the likelihood is also a non-logarithmic $\chi^2$. Similarly to what is done for the complete SED, when we give the code multiple data as input data, the final likelihood is also a sum of their reduced $\chi^2$, {each multiplied by the respective number of data points.}

The models of \beatlas cannot be directly compared to observational data: they must first be normalised for the distance of the object and corrected for the interstellar reddening. In our method, we correct for extinction using the prescription of \citet{fitzpatrick1999}, with both the colour excess, $E(B-V)$, and $R_V$ as free parameters. 




One of the reasons for choosing the Bayesian approach is its capacity to combine the data with prior knowledge. In principle any available information on a given target can be used as priors. Common examples are the parallax -- widely available thanks to the Gaia and Hipparcos missions (\citealt{gaia2016} and \citealt{perryman1997}, respectively) -- and the inclination angle that may be available from interferometric studies. Another important prior is $v \sin i$ that, however, should be applied with caution as it may become quite insensitive to the stellar rotation rate for large values of $W$ \citep{townsend2004}. In our study, we have the option to use any, none or all of these three priors {(more priors can be added to the code as necessity arises)}. We define the priors as Gaussian distributions centred in the literature value and with its literature uncertainty as variance. Therefore

\begin{equation}
\log p_a(\Theta,\alpha) \propto -0.5 \, \left( \frac{a_\mathrm{obs} - a_\mathrm{mod}}{\sigma_{a_\mathrm{obs} }} \right)^2\,,
\label{eq:prior_dist}
\end{equation}

\noindent where $a_\mathrm{mod}$ is the random model value {of parameter $a$} generated in each step of the inference. The priors used for each simulation shown here are specified in the text of Sects. \ref{sect:aeri} and \ref{sect:bcmi}.



\subsection{Simulation setup}

Summarising, the two grids defined in Sect.~\ref{sect:beatlas} have distinct applications. The photospheric grid can be used to study discless A {to} O stars, from non-rotating to fast rotators. Of great relevance is the study of inactive Be stars, for which \beatlas can directly probe the fundamental parameters without interference from the disc. In this case, the parameters explored by the MCMC simulations are: stellar mass ($M$), rotation rate ($W$), age (\ttms), inclination ($i$), distance ($\pi$) and extinction [$E(B-V)$]. In the case of active Be stars, the disc grid is used to explore additional parameters: disc base number density ($n_0$), disc radius ($R_\mathrm{D}$), and radial density exponent ($n$).

{For both cases, a set of derived parameters can also be recovered, as the stellar oblateness ($R_\mathrm{eq}$/$R_\mathrm{p}$), equatorial radius ($R_\mathrm{eq}$), the theoretical luminosity ($\log L$),  the mean effective temperature ($\overline{T}_\mathrm{eff}$), the gravity darkening exponent ($\beta_\mathrm{GD}$) and the projected rotational velocity ($v \sin i$). The stellar oblateness and $\beta_\mathrm{GD}$ are directly derived from the inferred $W$; $\log L$ is recovered from the Geneva grid, together with $R_\mathrm{p}$. $R_\mathrm{eq}$ is calculated using the stellar oblateness and $R_\mathrm{p}$. $\overline{T}_\mathrm{eff}$ represents an average value over the entire deformed Roche area, calculated through $\overline{T}_\mathrm{eff}^4 = L / (\sigma A_\mathrm{*})$, with $\sigma$ representing the Stefan-Boltzmann constant and $A_\mathrm{*}$ the Roche deformed area. Finally, $v \sin i$ is calculated using the inferred parameters $M$, $W$ and $i$ and the derived $R_\mathrm{p}$.}

A wide range of observables can be studied, including the SED from the UV up to the radio, continuum linear polarisation and hydrogen line profiles.


{There are already published works that used older versions of  \beatlas and our MCMC code \citep{dealmeida2020, suffak2020, marr2021, richardson2021, marr2022}, as they have been in development for over 5 years. However, the work we present here is the first critical test of our method, which has undergone major updates in the past year. Therefore, here we apply the method} to two well studied Be stars, \aeri (Sect.~\ref{sect:aeri}) and \bcmi (Sect.~\ref{sect:bcmi}), to certify that it can accurately recover the fundamental parameters of these stars. All of our simulations have 120 walkers in the ensemble. To ensure convergence, simulations were run until the number of steps reached 50 times the mean autocorrelation time of all parameters, as recommended by \citet{foreman2013}, and until the autocorrelation time of each parameter changed less than $1$ per cent between subsequent estimates. Typically, the above implies quite long chains with dozens of thousands of steps, depending on the dataset and grid used. In general, the simulations converge in a matter of hours or days, depending on the complexity of the model. 
To represent the PDFs obtained using the standard ``value $\pm$ uncertainty'' notation, we chose to use the highest density region (HDR -- also called highest probability region, or credible interval) of each PDF in $68$ per cent, with the {median} value of this region as the ``best'' value for the parameter \citep{hyndman1996, chen2000}. The HDR does not have to be unique and continuous, but can be broken in more than one region, which allows us to explore multiple solutions for non-unimodal distributions. Thus, when a distribution is double-peaked, two sets of ``solutions'' are considered by us as possible results. Furthermore, using HDR allows for a better representation of skewed distributions.



\section{\aeri: photospheric \beatlas} \label{sect:aeri}

\subsection{Target Overview}

\aeri ($\alpha$ Eridani, Achernar, HD10144, HR472, HIP7588) is the nearest Be star to Earth, at a distance of only $42.8 \pm 1.0$ pc \citep[$\pi = 23.39 \pm 0.57$ mas -- ][]{vanleeuwen2007}. It is well-known also for the drastic deformation it suffers due to its rapid rotation. Being such a close target, it has been the focus of several interferometric and imaging observations in the last decade. With PIONIER data, \citet{domiciano2014} imaged the photosphere of \aeri and measured its equatorial rotational velocity ($v_{\rm eq} = 298.8^{+6.9}_{-5.5}$
km s$^{-1}$), equatorial radius (\req $= 9.16 \pm 0.23 \, \mathrm{R_{\odot}}$), inclination
($i = 60.6^{+7.1}_{-3.9} \degree$), and the gravity darkening exponent ($\beta_{\rm GD} = 0.166^{+0.012}_{-0.010}$). The discovery of its wide binary companion, \aeri B \citep{kervella2007}, with a period of $2570.94 \pm 0.53$ days, $\approx 7$ years, and the subsequent {analysis of additional} observations of \citet{kervella2022} 
allowed for a robust estimate of the mass of \aeri A, the Be star, as $5.99 \pm 0.60\,\mathrm{M_{\odot}}$. It was also found to be an evolved star, with an estimated core chemical composition of $X \approx 6$ and $Y \approx 93$ per cent. 



The viscous disc of Be stars dissipates if matter stops being added to its base, i.e., if the mass ejection from the star ceases.
When this happens, the inactive
Be star is then perceived just as a fast rotating B star, as most of its class-defining observational traits disappear. \aeri has well-documented inactive phases (e.g., from 1970 to 1975 -- \citealt{vinicius2006}, and in early 2000s -- \citealt{rivinius2013b}), where no emission could be detected in its spectra, and several phases where a generally weak emission appears (e.g., between 1980 to 1990 and after 2002). Therefore, data from the inactive  periods can be used to study the  photospheric emission from \aeri.

\subsection{Results}




We applied the method described in Sect.~\ref{sect:methodology} for \aeri using the photospheric \beatlas grid. For the simulations presented here, we used a prior only on the parallax, using the Hipparcos value of $23.39 \pm 0.57$ mas. Note that there are no Gaia \citep{gaia2022} estimates for the parallax of \aeri as it heavily saturates the Gaia detectors. $R_V$ is kept fixed at the usual value of 3.1 \citep{fitzpatrick1999}.



The data used are UV spectra from the International Ultraviolet Explorer satellite (IUE -- \citealt{1976MmSAI..47..431M}), obtained from the INES Archive Data Center\footnote{\url{http://sdc.cab.inta-csic.es/cgi-ines/IUEdbsMY}}. The UV spectra are} from 1989 (a phase of weak disc emission). Only high-dispersion large-aperture IUE spectra were chosen, as they can be {correctly} flux calibrated. The data was also averaged and binned to match the resolution of the grid. 
To test the photospheric grid, we use a spectrum taken during one of the well-documented inactive phases.
The selected spectrum  was the cleanest from the period of inactive phase in 2000, from January 11 (Fig. \ref{fig:ha_aeri}), taken with ESO's FEROS (high-res echelle spectrograph, $R$ = 48000), mounted at the 1.52-m telescope in the La Silla Observatory. Unfortunately, this spectrum is still crowded with telluric lines and emission due to cosmic rays. The worst of these effects were removed from the spectrum using a sliding box outlier remover routine. The absorption lines \ion{C}{II} 6578.1 and 6582.9 are quite strong: the region affected was also removed from the spectrum for the simulation, as our models do not include carbon lines and could not reproduce this feature. The \halp spectrum is also binned to the resolution of the grid and normalised to velocity space. The systemic velocity is found by fitting an inverted Gaussian to the line wings. The velocity range considered for the analysis is $-1000$ to $1000$ km s$^{-1}$. 


\begin{figure}
    \centering
    \includegraphics[scale=0.8]{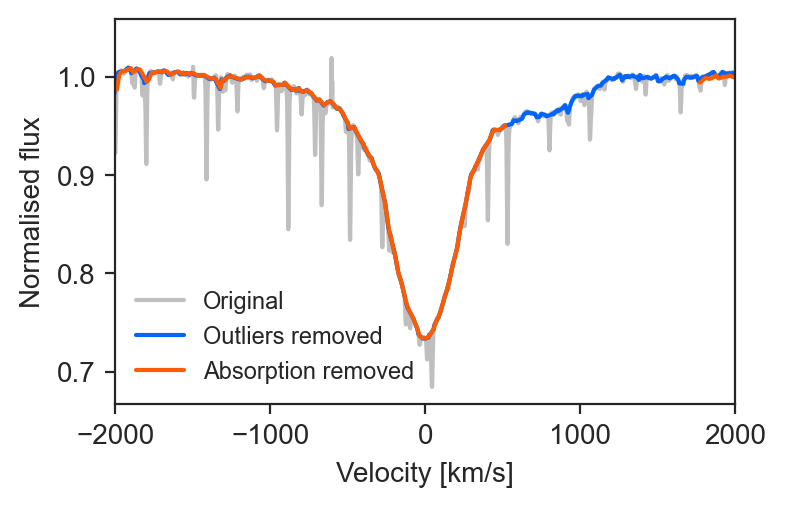}
    \caption{FEROS \halp profile for \aeri, from 2000 January 11, when the star was inactive. The original profile (grey line) was crowded with telluric lines and cosmic ray features, which were removed (blue line). The \ion{C}{II} absorption feature in ~6580\AA$\,$ was also removed (orange line).}
    \label{fig:ha_aeri}
\end{figure}

\begin{figure*}
    \centering
    \includegraphics[scale=0.5]{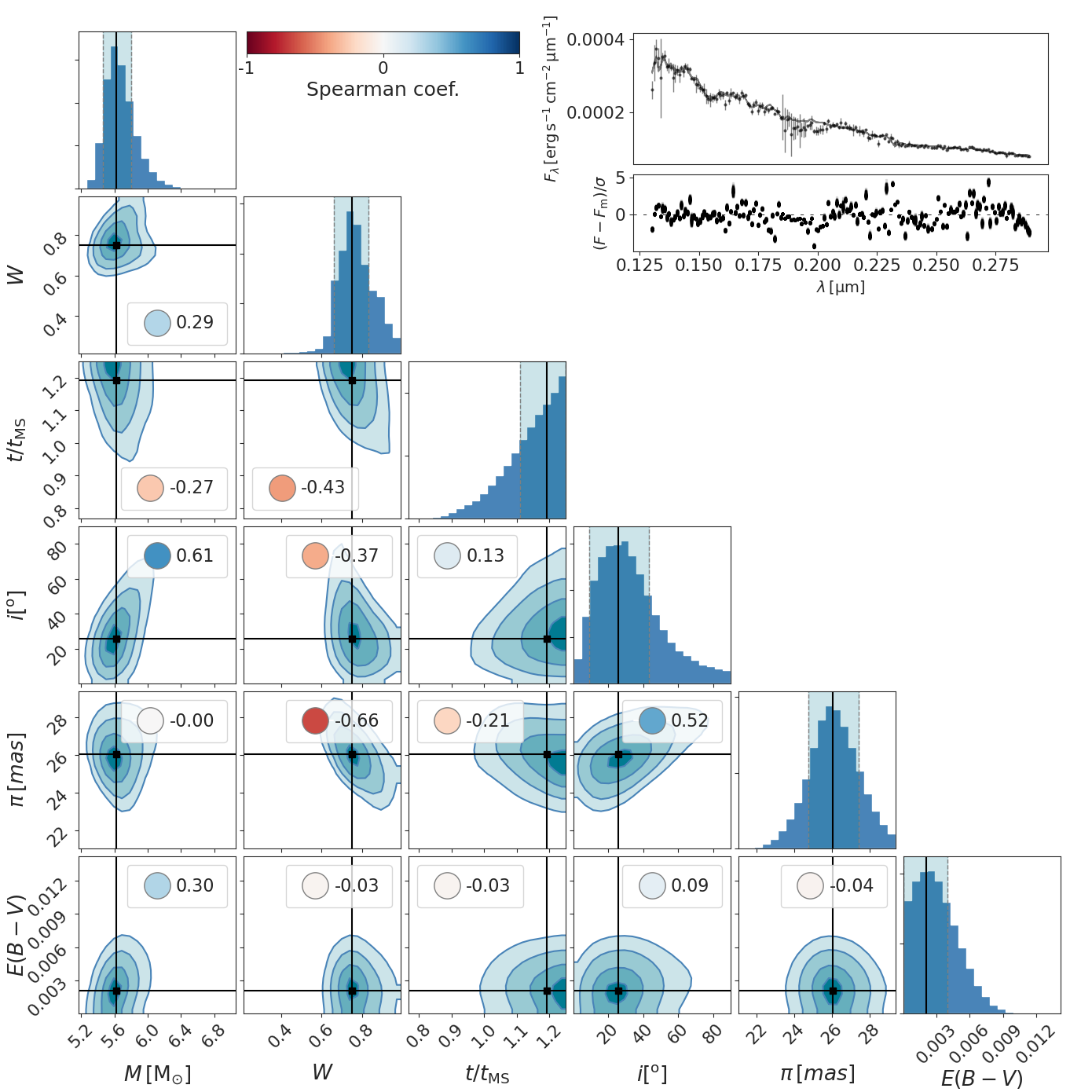}
    \caption{
  Corner plot of the UV-only simulation of \aeri. The PDFs for each parameter are shown in the main diagonal. The 2D histograms for each pair of parameters are shown in the lower part of the plot. The coloured circles indicate the {Spearman} coefficient for the correlation between the pairs of parameters (the colour scale are indicated at the top). On the upper inset, the observational data and residuals are plotted. The thin grey lines are a subset of 300 random models sampled by the code. 
    }
    \label{fig:aeriUV}
\end{figure*}

\begin{figure*}
    \centering
    \includegraphics[scale=0.5]{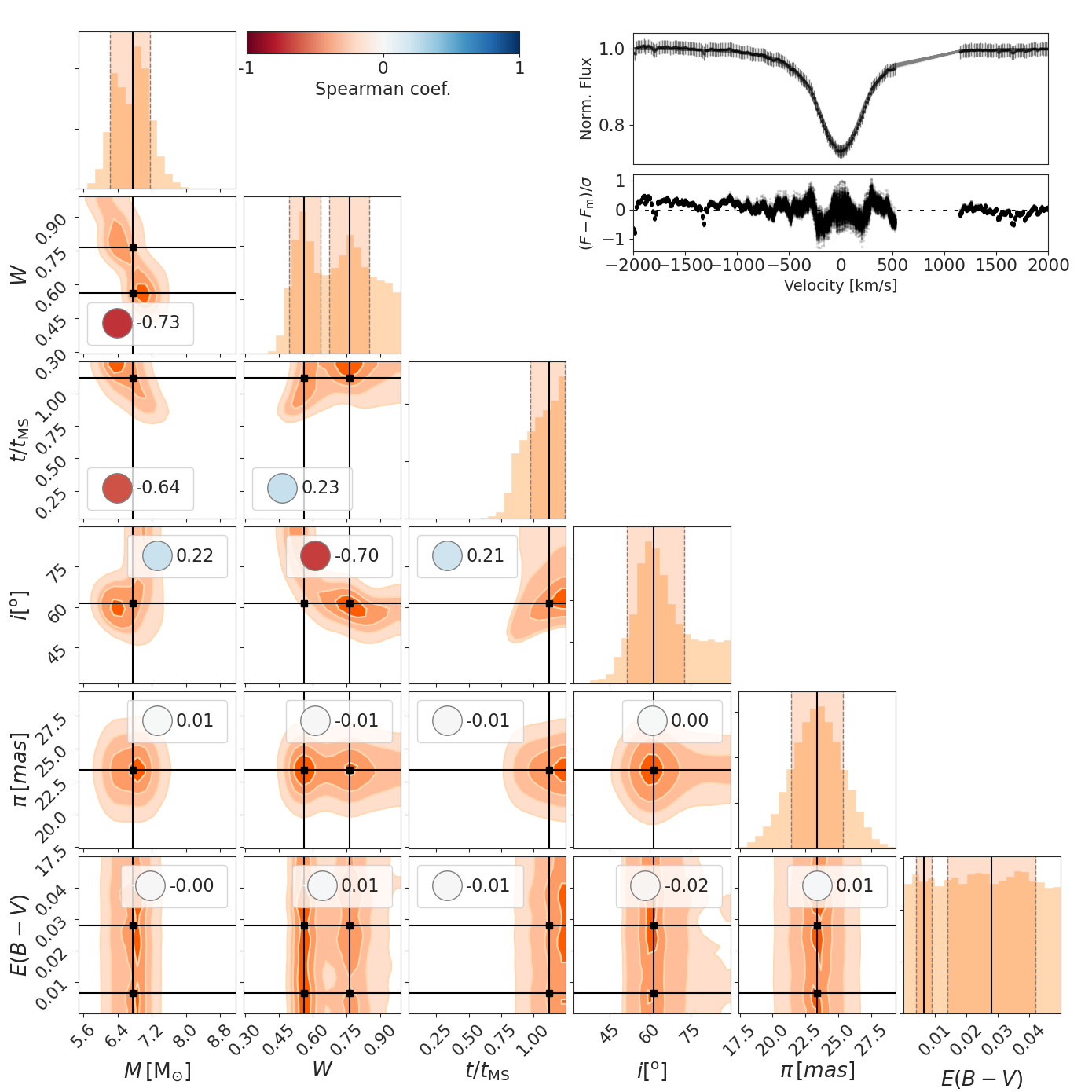}
    \caption{Same as Fig. \ref{fig:aeriUV}, but for the \halp profile of \aeri.}
    \label{fig:aeriHa}
\end{figure*}

\begin{figure*}
    \centering
    \includegraphics[scale=0.5]{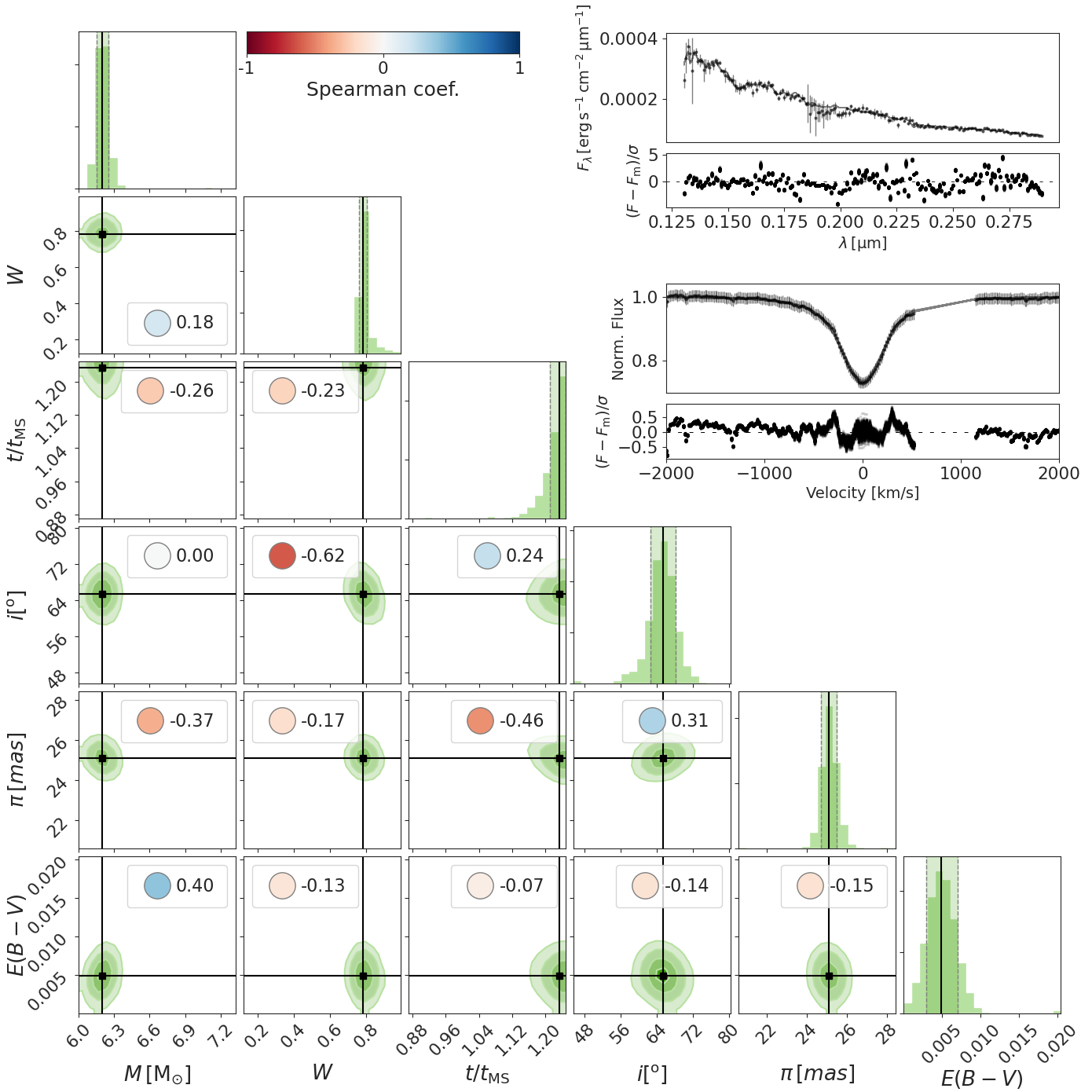}
    \caption{Same as Fig. \ref{fig:aeriUV}, but for both the UV spectra and \halp profile of \aeri.}
    \label{fig:aeriUVHa}
\end{figure*}

{To test our method and the constraining capabilities of different observables, } we ran three simulations in total, all with a prior on the parallax. For the first simulation only the UV data was used; the corner plot with the PDFs is shown in Fig.~\ref{fig:aeriUV}. On the second simulation, only the \halp profile was modelled; PDFs are shown in Fig.~\ref{fig:aeriHa}. Finally, we ran a simulation that considered both the UV and the \halp profile simultaneously (Fig.~\ref{fig:aeriUVHa}). {2D histograms and Spearman coefficients of all parameters for each simulation are also displayed in Figs.~\ref{fig:aeriUV}, \ref{fig:aeriHa} and \ref{fig:aeriUVHa}.} The results for all three simulations are summarised in Tab.~\ref{tab:aeri_res}. 

In Be stars, the UV and \halp emissions arise from completely different physical processes.
The UV comes mostly from the central star {(e.g., see Fig.~2 of \citealt{klement2015} and the work of \citealt{1978A&A....66..197B})}  and therefore is sensitive to the stellar parameters $M$, $W$ and \ttms. 
For fast rotators, it may also strongly depend on the inclination of the system. On top of that, the UV spectra is modified by the interstellar reddening and distance.


{The UV-only results (PDFs on Fig.~\ref{fig:aeriUV}) present \aeri as a $5.62_{-0.16}^{+0.18} \mathrm{M_{\odot}}$ Be star; the very well defined PDF for the mass agrees with the UV being sensitive to this parameter. They also bring \aeri closer than Hipparcos indicated, at 37.3 pc ($26.0_{-1.30}^{+1.36}$ mas), and tilt the system so the inclination is only about 25.9$\degree$, far from the interferometric estimate of 60.6$\degree$. We note that for both inclination and distance, the PDFs are broad, and thus the errors on these estimates are large: i.e., the {constraining power} of the UV on these parameters is not very strong. Even so, the large errors do not account for the discrepancy between these measurements, which means that the UV alone does not constrain the simulation enough to accurately estimate these quantities. Finally, a \ttms larger than one indicates that a star like \aeri is not well represented by the MS phase of the Geneva grid of evolutionary models. }



The \halp profile is sensitive to the stellar parameters  and, in the case of fast rotators, to the inclination, but not to the distance (as the profile is normalised) and reddening (as it has negligible effects in the line shape). The \halp-only results (PDFs on Fig. \ref{fig:aeriHa}) are consistent with these expectations, as the PDF for $E(B-V)$ is flat -- i.e., this parameter is undetermined by the simulation --, and the PDF for the distance simply reflects the prior used. The results indicates \aeri as a $6.75^{+0.42}_{-0.53} \mathrm{M_{\odot}}$ Be star, at the end of its MS lifetime, rotating at either $0.56^{+0.07}_{-0.06}$ or $0.76 \pm 0.09$ of critical. 

Comparing the UV- and \halp-only simulations shows that the only relatively well constrained stellar parameters are $M$ and \ttms -- albeit with large uncertainties. 
As discussed in Sect.~\ref{sect:photospheric_grid}, our grid allows for values of \ttms$ > 1$, i.e., the star has left the MS already (from the theoretical perspective of the Geneva models). The result from all our MCMC simulations place \aeri at this point in its evolution. {Our extrapolation of the Geneva models for the artificial \ttms$> 1$ ages is based on the native \hdust parameters actually used to calculate the radiative transfer models, $R_{\rm p}$ and luminosity. Therefore, our result of \aeri as an evolved star is indicative of its larger physical size and brightness rather than its actual age.}
The case of the rotation rate is interesting: it is the only parameter in our simulations to show a clear double-peaked PDF structure, with two separate HDRs (\halp-only simulation, Fig. \ref{fig:aeriHa}). The relative strengths of the two peaks indicate that both solutions are likely given the data. If we consider the highest $W$ found for the \halp, it is consistent with the UV-only result. The odd one out is the inclination: it is particularly discrepant between the two sets of results, with the UV-only result suggesting a lower value, inconsistent with previous studies of \aeri. However, the errors are large, as the UV does not constrain the parameter well. 
But all in all, given the limited data used in each simulation, the results are consistent with each other, {excluding the inclination}.


We cannot simply chose one set of results over the other. 
The MCMC method does not know stellar astrophysics; it is bounded by the data and the grid we give to it. 
What we can do to find the best possible results is to give the code more information to work with, so the parameters can be better constrained. This is where the result for the simulation using both UV and \halp shines (Fig.~\ref{fig:aeriUVHa}). {It is immediately visible that the PDFs for all parameters are thinner, and the uncertainties on the estimates of the parameters are thus smaller (Tab. \ref{tab:aeri_res}). The residuals in the inset plots also show the models fit both the UV and \halp line profile quite well, indicating that the simulation was successful in describing both data simultaneously.}

\begin{figure*}
    \centering
    \includegraphics[scale=0.9]{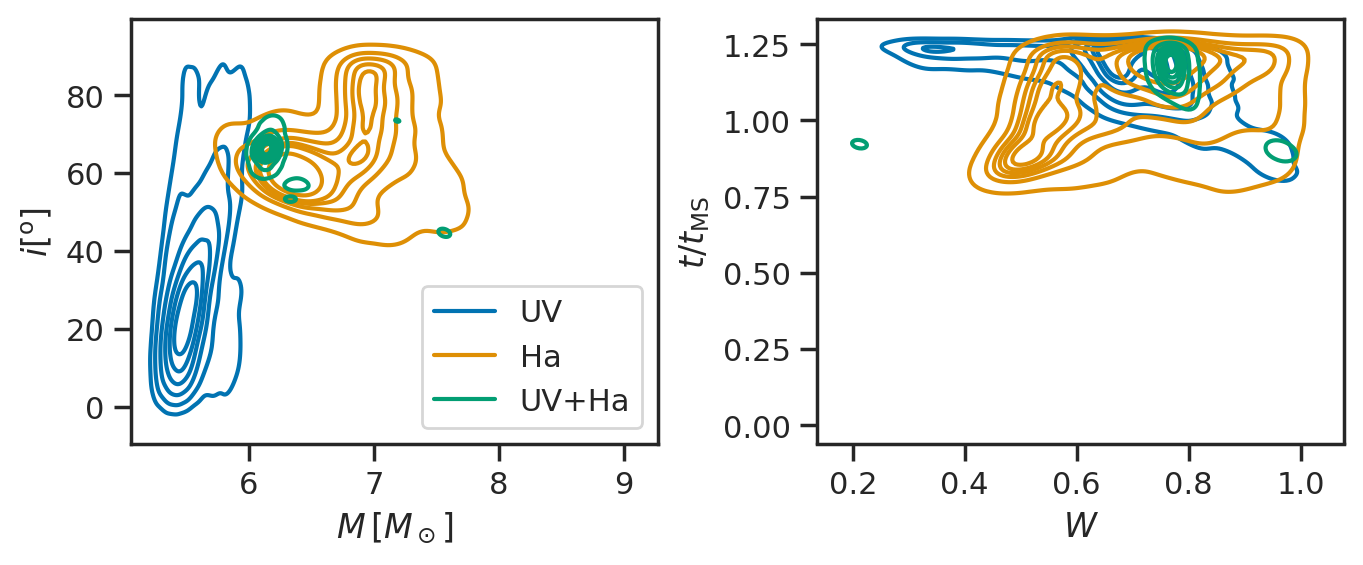}
    \caption{2D histogram for the mass and inclination and \ttms and $W$ of \aeri for our three simulations: UV (in blue), \halp (in orange), and UV and \halp combined (in green). Levels correspond to iso-proportions of the density for each distribution. 
    }
    \label{fig:aeri_compare}
\end{figure*}

When we consider both datasets at the same time, a tug-of-war between them is at play. Fig.~\ref{fig:aeri_compare} shows the 2D histogram for the mass and inclination (the two parameters where the UV- and \halp-only simulations are at odds) and for \ttms and rotation rate (where all models agree) for all three simulations. As shown by the contours, the UV- (in blue) and the \halp-only (in orange) results for $M$ and $i$ only overlap on a small region of the parameter space (in contrast with the \ttms and $W$ map, where the overlap region is much larger). When given both datasets at once, the UV+\halp simulation converges to the overlap region. For the rotation rate, the peak at $W = 0.56$ found on the \halp simulation disappears, in agreement with the fact that both \halp (considering the second peak) and the UV find a common solution closer to 0.77, which is astrophysically more significant as most known Be stars have rotation rates higher than 0.6 \citep{rivinius2013}. The mass is also very well defined by this combined simulation. The mass of a star is notably a difficult parameter to determine in general, and by far the best method for its determination is orbital analysis of visual binaries, where a $2$ per cent accuracy is the mass is a very precise result (for example, see \citealt{anguita2022}). The mass we determine with our method is more properly referred to as a ``theoretical'' mass, as it depends strongly on the underlying evolutionary model used in our methodology. 
This example corroborates the intuitive notion that more reliable and significant results should be obtained as more observations are included. 



{2D histograms and {Spearman} coefficients in Figs. \ref{fig:aeriUV}, \ref{fig:aeriHa} and \ref{fig:aeriUVHa} give us information on the correlations of each pair of parameter in the simulations. }
Positive {{Spearman} coefficients} indicate positive correlations: the simulation cannot discriminate well between increasing one or decreasing the other correlated parameters.
Negative {coefficients} (anti-correlations) are the opposite: increasing (or decreasing) one parameter or the other has a similar effect on the models. 
The parameters usually have correlations with more than one other parameter, which in turn correlates to others, creating a feedback that, given the large number of parameters involved, can become quite difficult to gauge. The untangling of these interconnections is one the main advantages of our method. In previous works, the correlations affecting the inference of stellar parameters of Be stars were not duly considered, as many parameters would be fixed to simplify the tiresome modelling procedure. 





\begin{figure}
    \centering
    \includegraphics[scale=0.65]{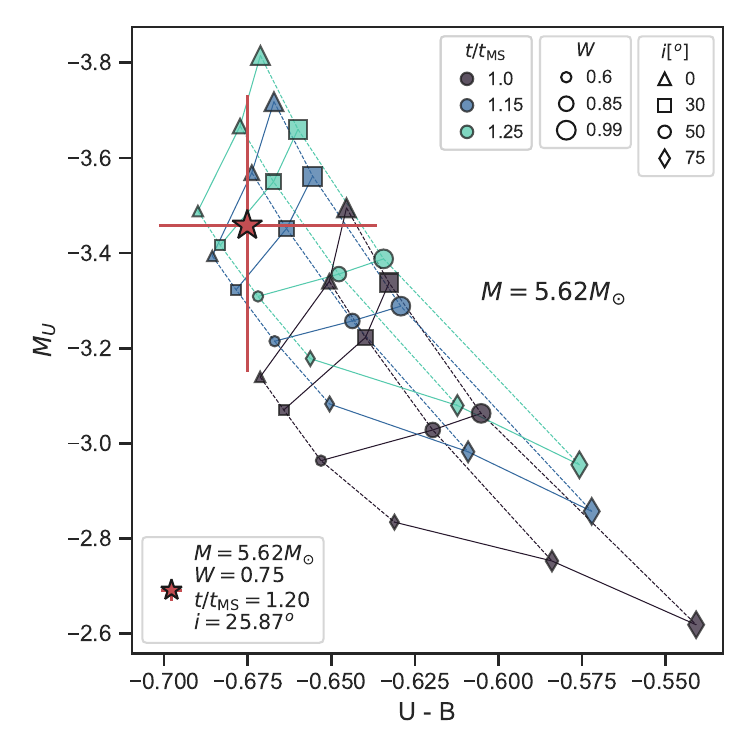}
    \caption{Colour-magnitude diagram for selected models from the Geneva grid. Different markers indicate inclinations, marker size indicate rotation rate $W$ and marker colour indicates \ttms. All models have a mass of $M = 5.62\, \mathrm{M_{\odot}}$ The red star marks the most likely solution given by the UV-only simulation for \aeri, as per Tab. \ref{tab:aeri_res}. }
    \label{fig:genebra}
\end{figure}

The UV-only simulation (Fig. \ref{fig:aeriUV}) shows relevant correlations between several parameters. 
The three stellar parameters are strongly intercorrelated, whereas $E(B-V)$ only correlates significantly with mass, and the distance only with $W$ and $i$. The inclination has relevant correlations to all stellar parameters and with distance, because changing the inclination of a Be star affects its apparent brightness given the oblateness and gravitational darkening of the star. Changing the distance also increases or decreases the apparent brightness of the star, which connects with $W$ and $i$ again because of the rotational effects on the surface gravity and area of the star. Therefore, we end up with a net of correlations between mass, rotation rate, \ttms, inclination and distance that is very difficult to disentangle. 

To help understand how the parameters correlate, we show in Fig. \ref{fig:genebra} the colour-magnitude diagram (CMD) of several models with $M = 5.62 \, \mathrm{M_{\odot}}$ (our result from the UV-only simulation), for different values of \ttms, rotation rate and inclination (see Fig. \ref{fig:full_genebra} for a version with several masses). The trends seen in the CMD for different families of models showcase the effects of each parameter. Increasing the inclination makes the star dimmer and redder, and increasing \ttms makes it brighter and bluer: they are positively correlated. However, changing the rotation rate can brighten or dim the star, depending on the inclination angle: for low inclinations, 
increasing the rotation rate leads to a brighter, but redder, star. Thus, as far as the UV flux is concerned, if $W$ increases, $i$ must decrease to keep the same general flux emission: they are anti-correlated. 
{The cluster of models with different $W$, \ttms and $i$ in areas of the CMD of Figs.~\ref{fig:genebra} and \ref{fig:full_genebra} show that there is significant degeneracies between the models, making it more difficult for the  simulation to discriminate between different parameters.}

The relationship between the parameters changes in the \halp-only simulation (Fig. \ref{fig:aeriHa}). While the inclination still correlates with the three stellar parameters, the distance and reddening {are not relevant for this simulation}. 
The rotation rate $W$ now has a strong anti-correlation ($\rho = -0.73$) with the mass, and a weaker positive correlation ($\rho = 0.23$) with the age. These correlations can be explained by the fact that the fast rotation of the star affects the shapes of the photospheric lines via rotational broadening, gravity darkening and a latitude dependent surface gravity. As such, the depth of the line decreases if the star rotates faster, but typically increases with decreasing mass (recall that the strongest H lines occur for A-type stars) and increases with decreasing surface gravity (or increasing radius).
The inclination and $W$ also have a significant anti-correlation ($\rho = -0.70$). The {gravity darkening} 
is more relevant if we can actually ``see'' the rotation: if the star is pole-on, for the observer it is not rotating at all, {and has a less extreme temperature gradient}. Therefore, if we decrease the inclination, the simulation pushes for higher rotations so its effects on the line profile are maintained. 

The correlation coefficients shown in Fig. \ref{fig:aeriUVHa} are the result of the combined influences of the UV spectrum and the \halp on the simulation. {The different values of the coefficients are} a reflection of their complicated relationships. 

\subsection{Discussion}

We now compare the results of the UV+\halp simulation for \aeri with the parameters obtained by \citet{domiciano2014} and \citet{kervella2022}, shown in Tab.~\ref{tab:aeri_res}.
The mass is remarkably consistent between our work and \citeauthor{kervella2022}'s (the mass was a fixed parameter in the analysis of \citealt{domiciano2014}), even though ours is a theoretical mass, derived from the Geneva models, and theirs is a dynamical mass, derived from interferometric and spectroscopic determinations of the orbital parameters of this binary system. \citet{kervella2022} also offers an estimate of the age of \aeri, in agreement with our results that point to a star either at the end of the MS or the early post-MS stages. Another remarkable result is the inclination angle, that agrees well with the interferometric determination of \citet{domiciano2014}.
The three estimates for the luminosity are similar within the errors and our derived distance is largely consistent with the Hipparcos value. The main disagreement is for $W$: our result ($W \approx 0.77$) suggests a lower rotation rate than found by \citeauthor{domiciano2014} ($W \approx 0.84$). Our estimate was constrained mostly by the \halp line profile, whereas \citeauthor{domiciano2014}'s by the interferometric measurement of the stellar oblateness. The different values of $W$ causes systematic differences between derived parameters such as $R_{\rm p}$, $R_{\rm eq}$ and $\beta_{\rm GD}$. Our estimate for the polar radius, $R_\mathrm{p} = 5.86^{+0.63}_{-0.55} \, \mathrm{R_\odot}$, coming from the Geneva models, is $\sim 14$ per cent smaller than \citeauthor{domiciano2014}'s estimate of $6.78 \, \mathrm{R_\odot}$. The smaller $R_\mathrm{p}$ and rotation rates lead to a large discrepancy between our estimate for the equatorial radius and \citeauthor{domiciano2014}'s.

The causes for these discrepancies between our results and literature lie with the underlying models, assumptions made in each analysis, {and the different data used}. The results of \citeauthor{domiciano2014} come from a geometric approach; by measuring the physical shape of the photosphere and the inclination of the rotation axis and assuming a rigid rotator model they obtain their value for $\beta_{\rm GD}$, and then derive $W$.
On the other hand, our model is spectroscopic: the shape of the line is what the code attempts to recreate.

{We note that both \beatlas and \citet{domiciano2014} use the rigid rotator approximation. This approach is common when modelling fast rotators, given that neither their interior nor their photosphere have a complete theory that covers all the physical phenomena observed in these stars. 
Therefore, discrepancies between models and data are possible. In the case of \aeri especially, its Be nature brings further complications. As our Fig.~\ref{fig:aeriHa} and Fig. 5 of \citet{domiciano2014} both show, there are features in the residuals when models are compared to the data. This can be due to physical phenomena that occur in the stellar photosphere that are not considered in the underlying models, or can also be caused by the presence of a weak, residual disc from \aeri's previous active phase. }

{Another factor that is not considered in our models is the fact that \aeri has a confirmed $2 \, \mathrm{M_{\odot}}$ companion. If the companion contributes significantly to the flux of the system, then a systematic error in the estimates of the parameters arises. This contamination is, however, negligible in the case of \aeri: the mean flux ratio is estimated to be around $2$ per cent \citep{kervella2022}, less than the uncertainties in most of our data. However, we point out that if this contamination by the secondary were indeed relevant, it would be trivial to sum its spectrum with the spectrum of the Be star when performing our MCMC simulations. }


All in all, our method's ability to recover the fundamental parameters of \aeri using solely the UV spectrum and a single \halp profile is a significant result, even more so as it surpasses previous works in speed and ease. Furthermore, with BeAtlas, the entire parameter space is explored at once, for all parameters, without the computational cost of needing to calculate new models for every Be star we wish to study. The results of our analysis showcases the reliability of \beatlas for studying inactive Be stars (as well as B and Bn stars), enabling seamless and semi-automatic modelling of numerous targets, with novel statistical robustness.




\renewcommand{\arraystretch}{1.5}
\begin{table*}
\begin{threeparttable}
\centering 
\caption{Results summary of our simulations for \aeri  and literature estimates of the same parameters. The first six listed parameters are the ones inferred directly from our simulations; the others are derived.}
\label{tab:aeri_res}
\begin{tabular}{llllll} 
\hline 
Parameter  & UV & \halp & UV + \halp & \citealt{domiciano2014} & \citealt{kervella2022} \\ 
\hline 
$M\,[\mathrm{M_\odot}]$& $5.62^{+0.18}_{-0.16}$ & $6.75^{+0.42}_{-0.53}$ & $6.20^{+0.05}_{-0.05}$ & $6.1^{\text{\ding{72}}}$ & $5.99 \pm 0.60$ \\ 
$W$& $0.75^{+0.08}_{-0.09}$ & $0.56^{+0.07}_{-0.06}$ or $0.76^{+0.09}_{-0.09}$ & $0.78^{+0.02}_{-0.02}$ & $0.84$ & --- \\ 
$t/t_\mathrm{MS}$& $1.19^{+0.06}_{-0.08}$ & $1.12^{+0.12}_{-0.15}$ & $1.24^{+0.02}_{-0.02}$ & --- &  $\approx 0.90$$^{\mathrm{(a)}}$\\ 
$i[\mathrm{^o}]$& $25.87^{+17.35}_{-16.74}$ & $61.29^{+11.50}_{-9.75}$ & $65.35^{+2.79}_{-2.75}$ & $60.6^{+7.1}_{-3.9}$ & ---$^{\mathrm{(b)}}$ \\ 
$\pi\,[\rm mas]$& $26.05^{+1.36}_{-1.30}$ & $23.39^{+1.98}_{-2.00}$ & $25.10^{+0.39}_{-0.39}$ & $23.39 \pm 0.57^{\text{\ding{72}}}$ & $23.39 \pm 0.57^{\text{\ding{72}}}$\\ 
$E(B-V)$& $0.002^{+0.002}_{-0.002}$ & $0.007^{+0.002}_{-0.003}$ & $0.005^{+0.002}_{-0.002}$ & 0.00 & 0.00 \\ 
$R_{\rm eq}/R_{\rm p}$ & $1.28^{+0.07}_{-0.06}$ & $1.16^{+0.04}_{-0.03}$ or $1.29^{+0.07}_{-0.06}$ & $1.31^{+0.02}_{-0.02}$ & $1.352$ & $1.39$\\ 
$R_{\rm eq}\,[\mathrm{R_\odot}]$ & $7.10^{+0.78}_{-0.72}$ & $7.05^{+1.08}_{-0.88}$ or $7.57^{+1.27}_{-1.06}$ & $7.68^{+0.21}_{-0.21}$ & $9.16 \pm 0.23$ & $8.14 \pm 0.26$$^{\mathrm{(c)}}$ \\ 
$\log(L)\,[\mathrm{L_\odot}]$ & $3.33^{+0.08}_{-0.09}$ & $3.60^{+0.17}_{-0.21}$ or $3.59^{+0.17}_{-0.21}$ & $3.50^{+0.02}_{-0.03}$ & $3.480$ & $3.54 \pm 0.05$ \\ 
$\beta_\mathrm{GD}$ & $0.181^{+0.013}_{-0.013}$ & $0.207^{+0.008}_{-0.010}$ or $0.178^{+0.013}_{-0.014}$ & $0.176^{+0.003}_{-0.004}$ & $0.166^{+0.012}_{-0.010}$ & --- \\ 
$v \sin i\,[\mathrm{km \, s}^{-1}]$ & $127.0^{+95.8}_{-86.3}$ & $210.1^{+56.2}_{-50.5}$ or $276.2^{+65.3}_{-65.0}$ & $278.8^{+14.5}_{-13.7}$ & $260.3$ & --- \\ 
$\overline{T}_{\rm eff}^{\mathrm{(d)}} [10^4 \, \mathrm{K}]$ & $1.546^{+0.059}_{-0.061}$ & $1.784^{+0.095}_{-0.135}$ or $1.741^{+0.113}_{-0.146}$ & $1.645^{+0.018}_{-0.021}$ & $1.5^{\text{\ding{72}}}$ & $1.5539 \pm 0.0438$ \\ 
\hline 
\end{tabular}
\begin{tablenotes}
      \small
      \item Notes. $^{\mathrm{(a)}}$The time fraction was estimated based on the same models used in Sect. 4.3 of \citealt{kervella2022}, from which the authors evaluated an age of $63 \pm 4$ Myr. $^{\mathrm{(b)}}$The inclination angle of $30.32 \pm 0.35$ calculated by \citealt{kervella2022} corresponds to the orbital plane of the system. The misalignment value between the equatorial plane of Achernar A and B of $30^{+7}_{-4}$ was calculated using $i$ from \citealt{domiciano2014}. $^{\mathrm{(c)}}$This value corresponds to a mean radius estimate fitted by \citealt{kervella2022}, and it's not exactly an equatorial radius measurement. {$^{\mathrm{(d)}}$In our work and in \citealt{domiciano2014}, $\overline{T}_{\rm eff}$ is an average over the Roche photosphere, while in \citealt{kervella2022} they find an average temperature from spectral fitting using \citet{castelli2003}}. \ding{72} Fixed value in their analysis.
    \end{tablenotes}
  \end{threeparttable}
\end{table*}

\section{\bcmi: disc \beatlas} \label{sect:bcmi}


\subsection{Target Overview}

\bcmi ($\beta$ Canis Minoris, HD58715, HR2845, HIP36188 -- B8Ve) 
is one of the most stable Be stars known, having been active since its discovery \citep{merrill1925}, maintaining a stable disc with little variation \citep{klement2015}. Also nearby \citep[$20.17 \pm 0.20$ mas,][]{vanleeuwen2007}, \bcmi was the target of interferometric studies: \citet{1997ApJ...479..477Q} placed a lower limit for the diameter of the \halp emitting region of $13 \, \mathrm{R_{\odot}}$, and  
later \citet{2005ApJ...624..359T}, also through long-baseline interferometric data, found the diameter to be $28.8\pm5.7 \, \mathrm{R_{\odot}}$.
\bcmi is the first late-type Be star to show signs of non-radial pulsation \citep{saio2007}; spectrointerferometry and spectroscopy analysis of \bcmi provided important confirmation of the Keplerian nature of Be discs \citep{Wheelwright2012,Kraus2012}.

\bcmi was the target of the multi-technique analysis of \citet{klement2015}. 
This work represents a landmark in the Be literature, as it is still the most comprehensive model of a Be star to date
\citep[note that a corrigendum was published,][]{klement2015corrigendum}. One interesting result was the detection of a SED turndown, a steepening in the slope of the SED in the radio, that was successfully modelled by a truncated disc, suggestive of a binary companion \citep[see also][]{klement2017}. {The authors also find the \ion{C}{IV} 1548 line in the spectra of \bcmi (not expected for a late-type B star), which they interpret as an indication of a hot, possibly subdwarf (sdB or sdO) companion}. \citet{dulaney2017} detected a companion in the system using radial velocity (RV) analysis, with a period of 170.4 days. The detection was, however, questioned by \citet{harmanec2019}, who noted that the RV signal might be contaminated. \citet{wang2018} also found no sign of a sdOB secondary when analysing \bcmi's UV data. Its status as a binary is therefore still not confirmed.



\subsection{Results}\label{sect:results_bcmi}



We test the disc \beatlas on \bcmi using virtually the same data as \citet{klement2017}, which is summarised below. Photometric data were combined from the literature using the Virtual Observatory VOSA \citep{2008yCat..34920277B}, comprising  results of several observers and missions: \textit{UBVRI} photometry from \cite{2002yCat.2237....0D}; IR photometry from \citet{dougherty1991}, IRAS \citep{cote1987}, AKARI/IRC mid-IR all-sky Survey \citep{ishihara2010}, SST \citep{su2006} and WISE \citep{cutri2014}; sub-mm and radio data from the VLA \citep{taylor1990}, JCMT \citep{waters1991}, IRAM \citep{wendker2000}, APEX/LABOCA \citep{klement2015}, CARMA \citep{klement2015} and JVLA \citep{klement2019}. For the UV, data from \textit{IUE} was used. As for \aeri, only high-dispersion large-aperture observations were selected and the data were averaged and binned to match the resolution of the grid. We also use polarimetric data taken from the Pico dos Dias Observatory, consisting of multi-epoch \textit{BVRI} linear photopolarimetry \citep[see][for more details]{klement2015}. {Our polarimetric data includes the dataset used in \citet{klement2015}, but has more recent points as well (Tab. \ref{tab:polarimetric_bcmi_data})}.



\bcmi can be considered a stable Be star, as its observables imply that the disc has not undergone dissipation or any significant structural changes \citep[such as $m = 1$ waves, which are common for Be stars - see][]{Okazaki1997} in the past decades. Fig.~\ref{fig:bcmi_bess} shows a compilation of 162 spectra from the BeSS database \citep{neiner2011} covering 20 years -- from 2002 to 2023. The data demonstrates the remarkable stability of \bcmi, displaying a root-mean-square (RMS) percentage variance of only $4.87$ per cent in the equivalent width (considering the median, $EW_{\rm med} = -3.77$
as the predicted value) and $2.79$ per cent in the V/R ratio (considering the predicted value as $\mathrm{V/R}=1$, also the value of the median of the measurements). 
Therefore, its SED, although comprised of data gathered in different epochs, is considered here as a true snapshot of its emission at a given moment. We also consider that the errors on any given measurement on the SED must be at least $10$ per cent of the measured value, as the intrinsic variability of a Be disc over time, even for discs as stable as this, should be accounted for. {As indicated by the RMS errors of \halp line measurements, $10$ per cent is a cautious and conservative estimate.} 
The polarisation data used for the simulation 
is an average of all measurements available for each of the 5 filters, and the errors are their median absolute deviation. 


Several simulations were calculated for \bcmi. We focus first on two of them: a full-SED simulation, with data
from UV to radio (0.13$\,\mu$m to 10$\,$cm) and a polarisation-only simulation that uses the same \textit{BVRI} polarisation data as \citet{klement2015}. We use a prior on the parallax for all simulations of \bcmi. {We use Hipparcos' determination, which has a smaller uncertainty and is well validated \citep{vanleeuwen2007}, rather than Gaia DR2's estimate.} 
For the full-SED simulation 
specifically, we also use a prior on the \vsini based on the value of \citet{2015ApJS..217...29B}: $248 \pm 13$ km s$^{-1}$. $R_V$ is kept fixed at the usual value of 3.1 \citep{fitzpatrick1999} for all simulations.




Fig. \ref{fig:bcmi_full} shows the result full-SED simulation for \bcmi. The PDFs for all parameters are very well defined, and the residuals indicate that the model was able to describe the data well. 
The picture of \bcmi our results paint is crystal clear: it is a late-type Be star, with an average rotation rate 
and at a later stage of its MS lifetime. 
The disc of \bcmi is on the low density end for Be stars at large, but at the high end for Be stars with similar spectral types -- see Fig.~8 of \citet{vieira2017}.
Interestingly, the results suggest a disc that is truncated at around $33\,$\req ($\approx 136\,\mathrm{R_{\odot}}$), and with radial density exponent $n$ smaller than 3.5, at about 2.8. 
Inclination, distance, and reddening are well constrained by the simulation.

As for \aeri, the correlations between parameters are complex, especially now that the disc is also being considered: as their non-negligible correlations indicate, stellar and disc parameters are coupled. Previous works 
for active Be stars do not take this into account, as they fix parameters (usually the stellar ones) when performing their $\chi^2$ {minimisation}. When {a parameter is fixed, its uncertainties are not propagated to the rest of the parameters, causing a chronic underestimation of the errors.} In addition, when correlations are ignored and the entire parameter space is not explored simultaneously, there is the danger of 
missing regions of the parameter space that {describe the system more accurately.} 

To understand the behaviour of the disc \beatlas simulation and the correlations of the parameters, we must understand how the disc affects the observables. The SED of a Be star originates from three main components: photospheric emission, disc emission (reprocessed from photospheric emission) and disc absorption. The disc of a Be star can be thought of as a pseudo-photosphere that grows with wavelength approximately as  \citep{vieira2016}

\begin{equation}
    R^{\rm eff}_{\lambda} \propto \lambda^{(u+2)/(2n - \beta - s)} \times \rho_0^{2/(2n - \beta - s)} \,, 
\end{equation}
where $s$ and $u$ are the exponents of the radial power laws that describe the temperature and Gaunt factor variations, $\beta$ is the exponent for the scale-height of the disc, as per Eq.~\ref{eq:scaleheight}, and $\rho_0$ is the base density of the disc, as per Eq.~\ref{eq:rhovdd}.
Thus, the size of the pseudo-photosphere relates to the wavelength of the emission, with larger areas being responsible for the emission in larger wavelengths (\citealt{carciofi2011}; see also Fig. 13 of \citealt{vieira2015}). The disc is then divided in two parts: an optically thick inner disc ($r < R^{\rm eff}_{\lambda}$) and a thin outer disc ($r > R^{\rm eff}_{\lambda}$), and the SED emission we measure is a combination of these. As such, we would expect the UV and visible to be more sensitive to flux from the star itself and the innermost disc, while the IR and the radio wavelengths (which originates from larger disc area) from the whole disc.

All three disc parameters are correlated, as all play a role in defining the pseudo-photosphere and thus the disc emission as whole. 
If $n$ increases, the density profile of the disc becomes steeper and the size of the pseudo-photosphere becomes smaller, which  translates in the SED as less flux excess. To remedy this effect, the disc would have to be denser. Indeed, this is the case for the full-SED simulation: $n_0$ and $n$ have the strongest positive correlation among all parameters ($\rho=0.78$).
The correlation of the disc size with the base density is also quite straightforward: if the disc is smaller, there is a decrease in the overall emission; increasing the base density would counterbalance this effect 
(the results in Fig.~\ref{fig:bcmi_full} indicate a weaker correlation, $\rho=0.38$).
Similarly for $R_{\rm D}$ and $n$, as the disc size decreases, the emitting area also decreases, leading to a drop in the flux excess. To balance this, $n$ would have to decrease as well to increase flux excess. In the full-SED simulation, the correlation is positive, as expected, but with a smaller significance ($\rho=0.29$).

The inclination of the system also greatly affects the emission that reaches the observer. For a Be star seen pole-on, the general effect of the disc is to increase the flux at longer wavelengths, as
reprocessed starlight is redirected mostly perpendicularly to the disc plane. 
For tenuous discs such as \bcmi's, this flux excess is almost negligible in the UV, but its significance increases with wavelength so that from the far IR forwards it completely dominates the SED. For edge-on stars, the disc obscures part of the star: if the pseudo-photosphere is small, it can dim the system as a whole, as the disc steals stellar flux, but does not have a large re-emitting area. On the other hand, if the pseudo-photosphere is large, the disc will emit more in redder wavelengths, now brightening the system. In the case of \bcmi, its intermediate inclination of $\sim$45$\degree$ means that the stellar flux is partially absorbed and reprocessed by the disc, but still contributes significantly to the SED emission. {In our simulation, we found the strongest correlations for $i$ were with the stellar parameters, while only relatively weak correlations exist with the disc parameters, indicating that the geometry and temperature of the star are more affected by the inclination than the geometry and emission of the disc.

As the disc, in some wavelength ranges, has an impact on the flux that can rival the star itself, 
there are correlations also between stellar and disc parameters. The mass does not correlate significantly with the disc parameters, but the age and $W$ do, in particular \ttms with $\log n_0$ and $n$, and $W$ with $R_\mathrm{D}$.
{The correlations with \ttms are straightforward. A larger \ttms implies a larger star and therefore a larger pseudo-photosphere for the same base density; the two parameters are then anti-correlated ($\rho=-0.43$ in Fig.~\ref{fig:bcmi_full}); the same principle is valid for the correlation between \ttms and $n$ ($\rho=-0.27$). The correlations between $W$ and the disc parameters also follow the same logic:} larger $W$ implies a larger star and larger pseudo-photosphere, an effect that requires $R_{\rm D}$ to decrease to be counterbalanced ($\rho=-0.16$ in Fig.~\ref{fig:bcmi_full}).




{The polarisation-only simulation is shown in Fig. \ref{fig:bcmi_pol}.}
The contrast between the results of the full-SED and polarisation-only simulations could not more striking. The {latter} simulation offers little to no constraint on the central star, but allows for estimating  the base density of the disc. This result is expected as the primary factor controlling the polarisation level, other than the geometry, is the total number of scatterers -- here, free electrons \citep{brown1977}. The polarisation-only simulation highlights a valuable lesson: without knowledge of the central star and on the density exponent $n$ (strongly correlated with $\log n_0$), the simulation was unable to accurately determine the disk parameters that {polarisation should be sensitive to. This results indicates that the constraining strength of the polarisation is too weak to distinguish between the models}. Thus, only the results from the non-flat PDFs of Fig. \ref{fig:bcmi_pol} were included in Tab. \ref{tab:bcmi_full}. Even so, we note that only the PDF for $\log n_0$ has enough statistical relevance. 

\begin{table*}
\begin{threeparttable}
\centering 
\caption{Results summary of our simulations for \bcmi {and literature estimates of the same parameters. The first nine listed parameters are the ones inferred directly from our simulations; the others are derived.}
}
\begin{tabular}{lllll} 
\hline 
Parameter & Full SED & Polarisation & \citealt{klement2015corrigendum} & \citealt{klement2017}\\ 
\hline 
$M  \, [\mathrm{M_\odot}]$& $3.64^{+0.02}_{-0.02}$ & --- & $3.5^{\text{\ding{72}}}$ & $3.5^{\text{\ding{72}}}$\\
$W$& $0.74^{+0.02}_{-0.02}$ &  --- & $\gtrsim 0.98$ & 1.0 \\
$t/t_\mathrm{MS}$& $0.71^{+0.02}_{-0.02}$ & $0.82^{+0.18}_{-0.31}$ & --- & ---\\ 
$\log \, n_0$ [cm$^{-3}$]& $12.17^{+0.03}_{-0.03}$ &  $11.83^{+0.31}_{-0.29}$ & $12.30$  & $12.22^{+0.30}_{-0.18}$ \\ 
$R_\mathrm{D} [\mathrm{R_\mathrm{eq}}] $& $33.66^{+0.93}_{-0.97}$ & --- & $35^{+10}_{-5}$  & $40^{+10}_{-5}$\\ 
$n$& $2.83^{+0.04}_{-0.04}$ & --- & 3.0 – 3.5 & $2.9 \pm 0.1$ \\ 
$i \,[\mathrm{^o}]$& $46.29^{+1.55}_{-1.54}$ &  $40.50^{+8.49}_{-9.35}$ & $43^{+3}_{-2}$ & $43^{\text{\ding{72}}}$ \\ 
$\pi [\rm mas]$& $20.18^{+0.20}_{-0.20}$ & --- & --- & --- \\ 
$E(B-V)$& $0.026^{+0.002}_{-0.002}$ & --- & --- & $0.01^{+0.02}_{-0.01}$ \\ 
$R_{\rm eq}/R_{\rm p}$ & $1.28^{+0.01}_{-0.01}$ &  --- & 1.49 & --- \\ 
$R_{\rm eq}\,[\mathrm{R_\odot}]$ & $4.12^{+0.15}_{-0.13}$ & --- & 4.17 & 4.20 \\ 
$\log(L)\,[\mathrm{L_\odot}]$ & $2.38^{+0.02}_{-0.02}$ & --- & $2.26$ & $2.26 \pm 0.01$ \\ 
$\beta_\mathrm{GD}$ & $0.181^{+0.003}_{-0.003}$ &  --- & $0.1367^{+0.0025}_{-0.0013}$  & --- \\ 
$v \sin i\,[\mathrm{km \, s}^{-1}]$ & $221.1^{+6.8}_{-7.3}$ & --- & 270 & --- \\ 
$T_{\rm eff, pole}^{\mathrm{(a)}} [10^4 \, \mathrm{K}]$ & $1.280^{+0.045}_{-0.037}$ &  --- & $1.3740$ & --- \\ 
\hline 
\end{tabular} 
\begin{tablenotes}
      \small
      \item {$^{\mathrm{(a)}}${As \citet{klement2015corrigendum} derives the effective temperature at the pole, the result shown in this table also corresponds to a $T_\mathrm{pole}$ estimate, based on Eq.~4.21 of \citet{cranmer1996}, rather than the typical average over the deformed Roche area.}}
    \end{tablenotes}
\label{tab:bcmi_full}
\end{threeparttable}
\end{table*}

\begin{figure*}
    \centering
    \includegraphics[scale=0.34]{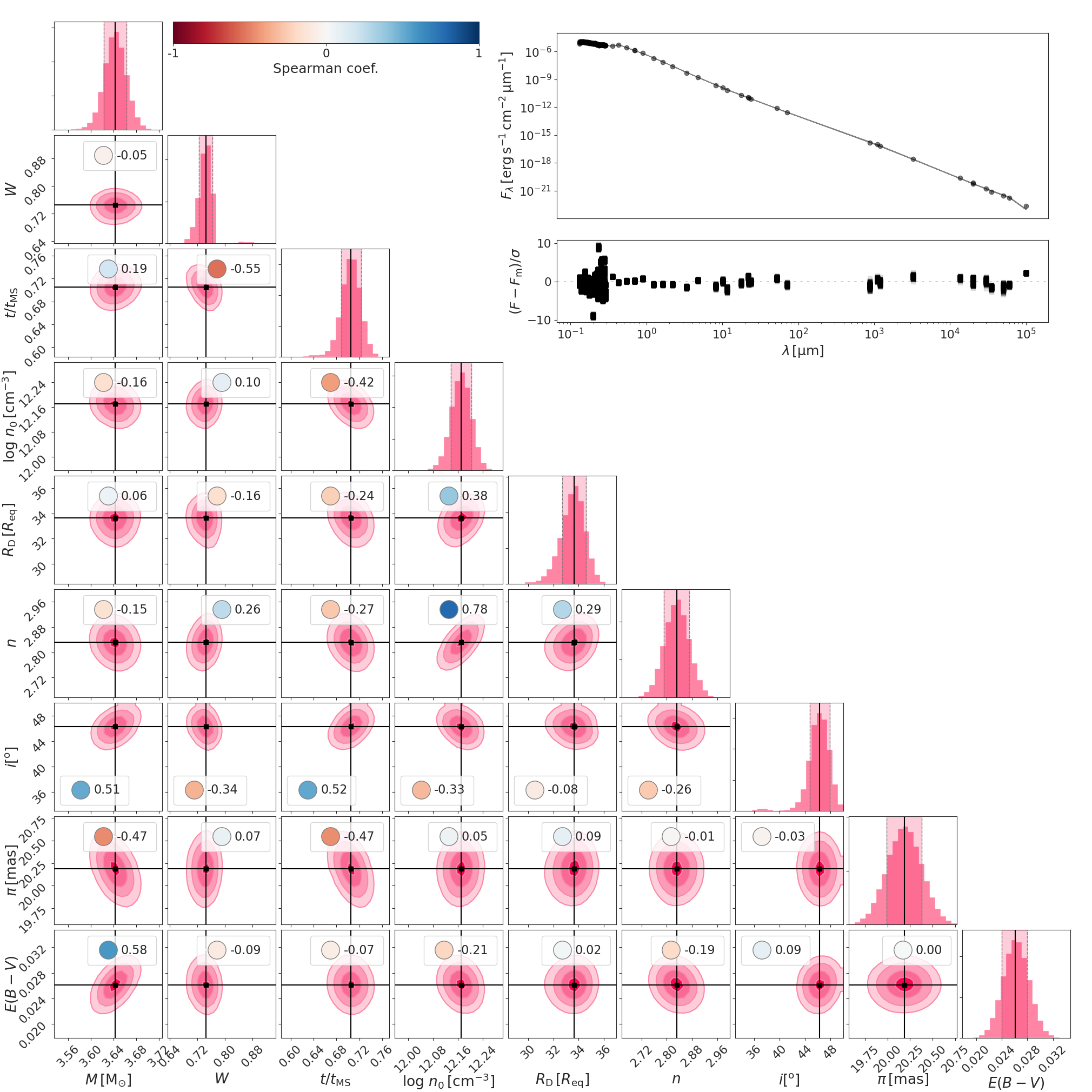}
    \caption{Corner plot of the MCMC run for the SED data of \bcmi. The PDFs for each parameter are shown in the main diagonal. The 2D histograms for each pair of parameters are shown in the lower part of the plot. The coloured circles indicate the {Spearman} coefficient for the correlation between the pairs of parameters. On the upper inset, the observational data and residuals are plotted. The thin grey lines are a subset of 300 of the models sampled by the code.}
    \label{fig:bcmi_full}
\end{figure*}

\subsubsection{Modelling Different Sections of the SED}\label{subsec:sections_sed}

If Be disc emission behaves as a pseudo-photosphere (Sect. \ref{sect:results_bcmi}), then this should be apparent in our simulations if only sections of the SED are given to the code as input rather than the full SED. 
To investigate this, and to study the constraining power of each part of the SED, we separate out data into two sections: one covering the UV, visible and near-IR (0.13 -- 5.0 $\mu$m) and another from the mid-IR to the radio (5.0 $\mu$m onward). 
The results are shown in Fig.~\ref{fig:box} in the form of violin plots. The different colours represent the three sections: UV-NIR (blue), MIR-Radio (orange) and full-SED (green). 
The corresponding corner plots of these simulations can be found in Fig.~\ref{fig:bcmi_uv-nir} and \ref{fig:bcmi_mir-radio}.
As expected, the PDFs of each parameter change drastically depending on the data used.
In the UV-NIR simulation, the stellar parameters are well defined, as expected, while the disc parameters are more scattered, showing broad, sometimes with double-peaked PDFs (blue violins). 
The opposite happens for the MIR-Radio simulation, where the PDFs for the stellar parameters are all flat, and the disc parameters are more defined (orange violins). 
The known correlations between both disc and stellar parameters (Sect.~\ref{sect:results_bcmi}) explain the different behaviours of these simulations. For instance, in the MIR-Radio simulation, the poorly constrained stellar parameters are reflected in the disc parameters. Here, a situation similar to the polarisation arises: although the MIR-Radio section of the SED should be the most effective in constraining the disc parameters, its ability to do so is hindered by the lack of information on the central star.



The differences between the results of the three simulations can also provide us with useful insights into structural changes in the disc.
The two discrepant estimates of $n$ by the UV-NIR and MIR-Radio sections, both very well determined by their respective simulations, can be a real effect, indicating that the inner disc is denser, with a less steep radial fall-off, than the outer disc, which has a larger $n$. 
{These results suggest that the radial slope of the density varies with the distance from the star, and the disc density is thus not well represented by a single power-law; in other words, it deviates from the isothermal steady-state VDD solution.}
This point will be further discussed below.

\begin{figure}
    \centering
    \includegraphics[scale=0.55]{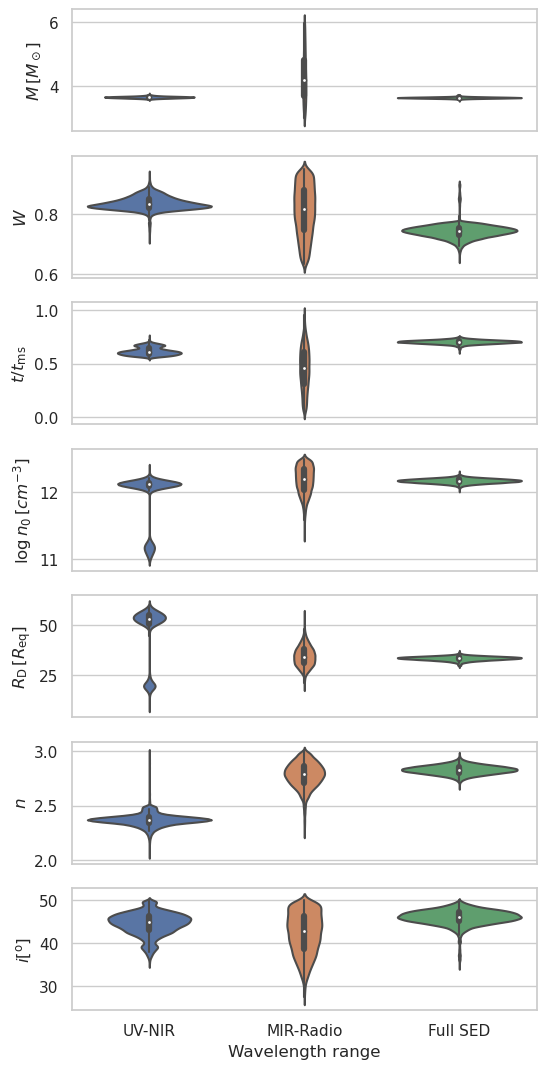}
    \caption{Violin plots showing the results of three MCMC simulations for different SED sections of \bcmi. The blue violins represent the PDFs of the parameters for the a section covering the UV, visible and near-IR (UV-NIR -- from 0.13 to 5.0 $\mu$m - corner plot in Fig.~\ref{fig:bcmi_uv-nir}); the orange violins for the section from mid-IR to the radio (MIR-Radio -- 5.0 $\mu$m onward - corner plot in Fig. \ref{fig:bcmi_mir-radio}); the green violins are for the simulation considering the entire SED, as also seen in Fig.~\ref{fig:bcmi_full}.}
    \label{fig:box}
\end{figure}

\subsection{Discussion}\label{sect:discussion:bcmi}


\bcmi was the target of an in-depth analysis by \citet{klement2015, klement2017}. They calculated a custom grid of \hdust models, keeping the mass and age of the star fixed. These models were compared to UV spectra (used to initially constrain the stellar radius and luminosity), photometric data ({according to \citealt{klement2015}, the IR data constrained the disc parameters}), polarisation ($W$ and $\beta$), AMBER spectrointerferometry ($i$), hydrogen line profiles (used for consistency checks) of \bcmi via $\chi^2$ minimisation. \citet{klement2017} include new radio data, but the same models were used for comparison. These works therefore provide a very complete view of the star, which made it a good target for testing our method. The abridged results from our simulations and from \citealt{klement2015corrigendum, klement2017} are on Tab. \ref{tab:bcmi_full}. 

The agreement between the results is remarkable, specially considering that our models explored more values for the parameters (mass and age especially), but used fewer data to constrain them. The disc parameters in particular all agree within 1$\sigma$, and the inclination and distance are also recovered. The result for $E(B-V)$ is also consistent with the estimate of $E(B-V) = 0.01$ from \citet{dougherty1994}. The largest discrepancy in between the results is for $W$, which can be explained by the fact that age has a significant correlation with $W$ ($\rho=−0.55$), and is not explored in \citeauthor{klement2015}'s grid as a free parameter.  


Using the SED and priors alone our method is able to recover the inclination of the system within 1$\sigma$ of the interferometric measurements, which is the most reliable way to estimate this parameter. The base density of the disc and the exponent $n$ are also consistent within 1$\sigma$. {This result is clear proof of what can be achieved with \beatlas: finding the fundamental parameters of both the star and the disc, simultaneously, went from being a 2+ years project where a bespoke grid of models had to be calculated and systematically compared to different data, to a ${\sim}1$ week effort}. It is also interesting to note that the result from the full-SED and the polarisation-only simulations also give compatible results for the base density, even though these observables probe different phenomena (free-bound recombination vs. electron scattering) and regions of the disc.

\subsubsection{The SED turndown {and denser inner disc} of \bcmi}\label{sect:sed_turndown}



As discussed in Sect. \ref{sect:vdd_model}, the power-law prescription of the steady-state, isothermal VDD is, although often used in Be literature, an approximation. Inconstant mass and AM feeding rates, {viscosity variations}, non-isothermal effects and binarity can all lead to density structures where the radial density exponent $n$ is no longer the theoretically predicted value of 3.5.

{The viscosity of VDDs is parametrised by the parameter $\alpha$. In the solution we presented in Sect. \ref{sect:vdd_model}, it is implied that this parameter is constant both in time and throughout the disc. However, the work of \citet{rimulo2018} indicates that $\alpha$ can vary, for instance, depending on the status of the disc, being larger during disc formation than during dissipation. Furthermore, \citet{ghoreyshi2021} finds evidence for an $\alpha$ variable also with radius for the Be star 28 CMa. Our results for the UV-NIR and MIR-Radio simulations (Fig.~\ref{fig:box}) suggest a smaller $n$ in the inner disc of \bcmi, which could be explained by a radially variable $\alpha$.}

Thermal effects in Be discs were studied by \citet{carciofi2008} by solving the viscous diffusion and energy balance in the VDD formalism. In this scenario, the effective slope of the disc ranges from 2 to 3 between $1 \lesssim r \lesssim 3$\,\req, and from 3 to 3.5 for $r \gtrsim 3$\,\req. Strong effects are only seen for very dense discs; for the less dense disc of \bcmi ($\log n_0 = 12.17^{+0.03}_{-0.03}$) it is likely that they, if present, would cause smaller changes in the density slope. {Therefore, isothermal effects may also be the cause of the radial variation in $n$ we detect for \bcmi. } 

\citet{klement2015} reported a steepening of the slope of the SED in the radio region for \bcmi. The putative cause of this SED turndown is disc truncation: beyond a certain distance from the star, at the truncation radius, there is a sudden decrease in disc density. As the SED emission is dependant on the area of the emitting region, this density break translates as less flux in longer wavelengths. Our results (Fig.~\ref{fig:bcmi_full}, Tab.~ \ref{tab:bcmi_full}) indicate that the disc of \bcmi is truncated at about 34 \req, or about 139 $\mathrm{R_{\odot}}$ assuming \req = 4.1 $\mathrm{R_{\odot}}$, in agreement with the values reported by \citet{klement2015, klement2017} within 1$\sigma$. The physical cause of disc truncation is not clear. In the context of the VDD model, there are two likely causes: the presence of a {close binary companion and disc evaporation by gas pressure (disc ablation was believed to be another possible cause of truncation in the last decade, but more recent works -- \citealt{kee2016} and following series -- indicate its effect is much more significant in the inner disc than the outer disc.}

The radial velocity of a VDD grows with radius, starting with subsonic velocities at the base of the disc \citep{bjorkman2005}.
In principle, a VDD in an isolated (single) Be star should expand indefinitely, ultimately merging with the interstellar medium.
\citet{okazaki2001} showed that once the radial speed of the disc reaches the transonic regime (at the transonic critical radius), the azimuthal velocity is no longer Keplerian, but AM-conserving. This shift occurs because the
disc is no longer driven by viscosity, but rather by gas pressure. The critical radius relates to $v_{\rm orb}$ and the sound speed $c_s$ as \citep{krticka2011}
\begin{equation}
    \frac{R_s}{R_{\rm eq}} =  \left[ \frac{3}{10 + 4s} \left( \frac{v_{\rm orb}}{c_s} \right)^2 \right]^{\frac{1}{1 - s}},
\end{equation}
where $s$ is the power-law exponent for the temperature, $T = T_0 (R_{\rm eq}/r)^s$. For an isothermal disc, $s = 0$ and the typical values for $R_s$ are about 430 \req for a B9V star and 350 \req for a B0V; for \bcmi, using our most likely values, $\sim 315$ \req, {much further than our estimate for the truncation radius of \bcmi and other Be stars}. {However, the recent work of \citet{cure2022} revisits this idea and offers a different solution to the equation of motion. They find that for certain values of line-force parameters and disc temperature, the transonic point can occur much closer to the Be star, at radii smaller than 50 \req. This line of research will be further developed, but it is a possible answer to the truncation of Be discs.}




We are then left with the presence of a binary companion as the most likely cause of disc truncation for \bcmi. {As stated in Sect.~\ref{sect:central_star}, binarity can play a significant role as a spin-up channel of Be stars, supported by the existence of a large (and increasing) number of confirmed Be+NS and Be+sdO binaries.} \citet{pols1991} concluded that no more than $60$ per cent of the total number of Be stars would be formed this way, while \citet{vanbever1997} found a maximum of $20$ (and a minimum of $5$) per cent.

Several recent works, however, put much more weight in favour of the binary channel. Be stars have a higher fraction of runaways than normal B stars \citep{boubert2018} and are not found in systems with a low mass MS star \citep{bodensteiner2020}. The results of \citet{McSwain2005} photometric survey showed that up to $73$ per cent of Be stars may be a result of binary mass transfer. \citet{demink2013} indicated that the main channel through which massive stars in general are spun-up is binary mass transfer and mergers, and noted that their models produce faster spinning stars than the ones of \citet{pols1991} and \citet{vanbever1997}. The fraction of mergers and mass gainers in their simulation was $24.1$ per cent, very close to the estimate of the number of Be stars among early-type, non-supergiant B stars ($20$ -- $30$ per cent - \citealt{zorec1997}). \citet{shao2014} performed a thorough study of the influence of binarity in the birth of Be stars in the Galaxy using population synthesis, finding that fraction of Be stars that are a direct result of binary evolution among B-type stars was around $13$ – $30$ per cent, again remarkably close to \citet{zorec1997}'s estimate for the Be population among early B-types. {Recently, \citet{elbadry2022} found a progenitor of a Be + stripped star system, indicating that $10$ -- $60$ per cent of Be stars go through mass transfer and now have a sdO/B as companion.} It is interesting to note that the only Be star with convincing data \textit{against} the spin-up via binary mass transfer scenario is Achernar \citep{kervella2022}. 

Of particular interest is the work of \citet{klement2019}, which analysed the SEDs of 57 Be stars in search of signs of SED turndown.
The authors detected it for all 26 targets with sufficient data coverage (including \bcmi); among them are both confirmed and unknown binaries. Thus, the binary fraction among Be star could be as high as $100$ per cent if binarity is indeed behind the detected disc truncation.

Assuming a circular, coplanar orbit for the companion, \citet{panoglou2016} finds in their SPH simulations that the truncation radius of the disc has a 3:1 resonance with the orbit for binary Be stars. {No companion was detected for \bcmi in the UV \citep{wang2018}, nor via interferometry (Klement et al., in prep). Summing that to what is implied by the spectroscopic analysis of \bcmi \citep{dulaney2017, harmanec2019}, the mass of the companion must be low.} Thus, assuming a mass range for the possible companion of $0.5 > M_2 > 0.1 \, \mathrm{M_{\odot}}$, our derived truncation radius and that truncation happens at the 3:1 resonance radius, this would place orbital period of the secondary between 280 and 295 days. \citet{dulaney2017} found a possibly significant period at 170.4 days in their analysis of radial velocity shifts in \bcmi spectral lines. It does not match the 3:1 resonance, but lies somewhere between 3:2 and 2:1 resonances if masses below $1.0 \, \mathrm{M_{\odot}}$ are considered, and matches 2:1 for $M_2 \approx 1.3 \, \mathrm{M_{\odot}}$. Assuming our upper ranges for the values of $R_\mathrm{D}$ and \req, the period of 170.4 days is consistent with a 3:2 resonance for a companion of mass $M_2 \approx 0.1 \, \mathrm{M_{\odot}}$. We cannot discard the possibility of a possible companion as proposed by \citet{dulaney2017}, but their derived period and mass of the companion ($0.42 \, \mathrm{M_{\odot}}$) together cannot be reconciled with our truncation radius in terms of its relation with the resonance radius. This relation, however, is also dependant on factors not considered in our analysis, such as the viscosity of the disc and the eccentricity and misalignment of the orbit.

Another effect of binarity, discussed briefly in Sect.~\ref{subsec:deviations},
is the alteration of the radial density slope of the disc.
This change is due to the accumulation of matter that occurs as the tidal interaction with the companion star hinders disc growth, resulting in a denser inner disc.
 How intense the accumulation effect is depends on the separation of the two stars, their mass ratio and the disc viscosity: small orbital separation, large mass ratio and lower viscosity lead to the densest Be discs, with $n$ dropping to values as low as 2.5 \citep{panoglou2016}. 
 
Thus, the scenario depicted in Sect. \ref{subsec:sections_sed} (see Fig.~\ref{fig:box}), where the slope of the disk density ($n$) increases with distance from \bcmi, is consistent with the predicted accumulation effect observed in SPH simulations and could be interpreted as an indirect evidence of binarity.
However, as mentioned earlier, non-isothermal effects may also contribute to similar behaviour, making it challenging to differentiate between the two and determine their relative importance.

{If a companion is indeed present in the system, its influence would go beyond the deformation of the disc. A companion would contribute to the flux emission, particularly in the case of a hot sdOB companion. If this is the case, it also follows that the two stars likely evolved together and went through a period of mass transfer. As binary evolution is not considered in the underlying evolutionary models used in \beatlas, further systematic errors can be introduced in our inference of the parameters of the Be star. Furthermore, the companion could also illuminate and heat up the outer disc \citep{peters2008}, affecting the estimates of the disc parameters. All these effects are not accounted for in our work nor \citeauthor{klement2015corrigendum}'s models.}

\section{Conclusions}\label{sect:conclusions}





To fully understand a star, we require knowledge of its photospheric parameters, including $T_\mathrm{eff}$, $\log g$, and others. However, determining these parameters is a complex process that typically involves comparing theoretical models, such as synthetic spectra from model atmosphere calculations, with observations. The analysis is even more challenging for classical Be stars due to their rapid rotation and the presence of their circumstellar viscous Keplerian discs (as discussed in Sect.~\ref{sect:vdd_model}).


In this work, we present \beatlas, a grid of synthetic spectra of Be stars. It comprises two systematic grids of models: a purely-photospheric (i.e., discless) grid, and a star + disc grid (``disc grid''), computed according to the VDD formalism. The grid is composed of $616\,000$ synthetic spectra, calculated using modern stellar evolution models coupled with three-dimensional NLTE radiative transfer calculations. \beatlas integrates a significant amount of the knowledge accumulated in recent years regarding Be star discs.
The grid is used in conjunction with Bayesian MCMC sampling for the determination of fundamental parameters of the star, such as mass, rotation rate, age, and inclination. It also helps determine the properties of the disc, such as disc size, density scale, and density slope, as well as the distance and interstellar reddening of the target.
This method efficiently provides PDFs for the modelled parameters and their cross-correlations, even in multi-dimensional parameter spaces. A key advantage of this approach is the use of prior information about the object, which contributes to the inference process.

The power of the combination of Bayesian inference with the \beatlas grid is made quite clear by our results. As a first test of \beatlas, we demonstrate that it can be used to perform multi-technique analysis of both active and inactive Be stars. This was done by a detailed study of the active late-type Be star \bcmi and an inactive phase of late-type \aeri. 
These targets were selected because precise determinations for their fundamental parameters are available in the literature.
{We able to recover literature determinations of most stellar and disc parameters for our targets with good precision. The mass and age of \aeri both agree with the results of \citet{kervella2022}, which used interferometric and spectroscopic data of this binary system. The results for \bcmi are notable: with only the SED data as input to our simulation, a complete picture of the star and its disc is recovered, closely matching literature results. For both targets, the distance and inclination are also recovered; for \aeri, the inclination agrees with interferometric measurements. The biggest discrepancies between our results and the literature lie with the rotation rates of both stars, which are less dramatic than the previously reported $>98$ per cent of critical for \bcmi (our result is closer to $75$ per cent) and $84$ per cent for \aeri (against our result of $77$ per cent). For \bcmi in particular, we recover the disc truncation result of \citet{klement2015, klement2017}, a sign of {possible} binarity for Be stars. By considering separate sections of the SED in our MCMC sampling, we also find signs that the inner disc of the star is denser than the outer disc, another indicator of binarity. Thus, \bcmi is likely in a binary system; whether the companion is indeed the responsible for the signal detected by \citet{dulaney2017} is not confirmed. }

The correlations between all parameters were explored in unprecedented fashion. The net of inter-correlations is complex because Be stars are fast rotators: as evidenced in Figs.~\ref{fig:genebra} and \ref{fig:full_genebra}, adding rotation as a parameter leads to a multiple degeneracies between the stellar parameters. When we add the disc, whose emission is dependent on its density structure, wavelength considered and observing angle, the true complexity of Be stars is unveiled: if we want to properly study a Be star, the fact that the star and disc are coupled must be taken into account, which our method does seamlessly.




The applications of \beatlas are numerous. It can be used effectively to find estimates for the fundamental parameters of any given Be or B star with just the SED and \halp line profile, and adding more observables in the simulation (other H lines and polarisation, for instance), will likely increase the accuracy in which the parameters are recovered. \beatlas can also be used as a discovery tool: as our method allows for many different combinations of observables and priors, we are able to explore different emitting areas of the disc with more detail, building a more realistic image of Be star discs. 

{One key parameter of VDDs that is not explored in our grid is the viscosity parameter $\alpha$. As briefly discussed in Sect. \ref{sect:sed_turndown}, recent works have shown that $\alpha$ could be variable in time and with radial distance. In the prescription of \beatlas, these variations of $\alpha$ would translate into variations in our disc parameters (base density, density exponent and disc radius), meaning that, indirectly, our method accounts for these changes. The current version of the disc grid could be used as a diagnostic tool for $\alpha$ variations if a particular target had well-documented phases of disc build-up and dissipation: analysing these snapshots of evolution separately with \beatlas could give insight into the disc physics to uncover information on the viscosity and its variation. }

Since a model of a Be star can now be made in a matter of hours -- as opposed to weeks or months as previously  -- \beatlas can also be effectively used to study populations of B and Be stars. Comparing the characteristics of the two populations can reveal the necessary ingredients for the Be phenomenon. {In conclusion, we believe \beatlas has the power to massively upgrade the field of Be stars and help us solve our most pressing questions about these objects.}

\section*{Acknowledgements}

This study was financed in part by the ``Conselho Nacional de Desenvolvimento Cient\'{i}fico e Tecnol\' ogico'' - Brasil (CNPq) - Finance Code 140171/2015-0.
A.\,C.\,R. acknowledges support from FAPESP (grant 2017/08001-7), CAPES (grant 88887.464563/2019-00), DAAD (grant 57552338) and ESO. A.\,C.\,C. acknowledges support from CNPq (grant 311446/2019-1) and FAPESP (grants 2018/04055-8 and 2019/13354-1). P.\,T.\, acknowledges support from CAPES (grant 88887.604774/2021-00). 
M.G. acknowledges support from FAPESP (grant 2018/05326-5).
This work made use of the computing facilities of the Laboratory of Astroinformatics (IAG/USP, NAT/Unicsul), whose purchase was made possible by the Brazilian agency FAPESP (grant 2009/54006-4) and the \mbox{INCT-A}. 
CEJ acknowledges support through the Natural Science and Engineering Research Council of Canada (NSERC).
T.H.A. acknowledges support form FAPESP (grants 2018/26380-8 and 2021/01891-2).
C.A. and M.C. thanks the support from Centro de Astrofísica de Valparaíso. M.C., C.A. and I.A. acknowledge the support of Fondecyt projects 1190485 and 1230131 and ANID-FAPESP project 2019/13354-1.

This work has been possible thanks to the use of AWS-U.ChileNLHPC credits. 
{This work was also based on INES data from the IUE satellite.}

Powered@NLHPC: This work was partially supported by the supercomputing infrastructure of the \href{https://www.nlhpc.cl/}{NLHPC}\footnote{\url{nlhpc.cl}} (ECM-02). This work was performed using HPC resources from the computing centre \href{mesocentre.centralesupelec.fr/}{Mésocentre}\footnote{\url{mesocentre.centralesupelec.fr}} of CentraleSupélec and École Normale Supérieure Paris-Saclay supported by CNRS and Région Île-de-France. 
This research was enabled in part by support provided by \href{sharcnet.ca}{SHARCNET}\footnote{\url{sharcnet.ca}} and the \href{alliancecan.ca}{Digital Research Alliance of Canada}\footnote{\url{alliancecan.ca}}.

This study was granted access to and greatly benefited from the HPC resources of SIGAMM infrastructure (cluster Licallo), hosted by Observatoire de la Côte d’Azur (\url{crimson.oca.eu}) and supported by the Provence-Alpes Côte d’Azur region, France.

The authors acknowledge the National Laboratory for Scientific Computing (LNCC/MCTI, Brazil) for providing HPC resources of the \href{sdumont.lncc.br}{SDumont}\footnote{\url{sdumont.lncc.br}} supercomputer, which have contributed to the research results reported within this paper.

\section*{Data Availability}
The data underlying this article are available in the article and in its online supplementary material.
\bibliographystyle{mnras} 
\bibliography{Bibliography}

\begin{appendix}
\onecolumn
\newpage
\section{\textsc{BeAtlas}' Observables}\label{app:beatlas_observables}

\begin{table}
\centering
\begin{threeparttable}
\caption{Definition of \beatlas' observables.}
\label{tab:beatlas_observables}
\begin{tabular}{lccccccl}
\hline \hline
Observable & $\lambda_{\mathrm{min}}\,[\mathrm{\mu m}]$ & $\lambda_{\mathrm{max}}\,[\mathrm{\mu m}]$ & $N_{\mathrm{bins}}$ & Spacing & $N_{\mathrm{phot}}$ & Image & Comments\\ \hline
UV & 0.1 & 0.32 & 210 & linear & $4.0 \times 10^8$ & No & Kurucz resolution \\
SED & 0.32 & 1.05 & 108 & log & $1.0 \times 10^9$ & No & --\\
J & 1.05 & 1.4 & 30 & linear & $1.8 \times 10^8$ & No & --\\
H & 1.4 & 1.85 & 30 & linear & $1.5 \times 10^8$ & Yes & PIONIER$^{\mathrm{1}}$, MIRC$^{\mathrm{2}}$, R$\sim$300\\
K & 1.85 & 2.45 & 30 & linear & $1.5 \times 10^8$ & Yes & GRAVITY$^{\mathrm{3}}$, R$\sim$4,000\\
L & 2.45 & 3.9 & 16 & linear &  $1.0 \times 10^8$ & Yes & MATISSE$^{\mathrm{4}}$, R$\sim$1,000\\
M & 3.9 & 8 & 24 & linear & $1.5 \times 10^8$ & Yes & MATISSE, R$\sim$550\\
N & 8 & 13 & 15 & linear & $1.0 \times 10^8$ & Yes & MATISSE, R$\sim$250\\
Q1 & 13 & 18 & 10 & linear & $2.0 \times 10^7$ & No & --\\
Q2 & 18 & 25 & 10 & linear & $2.0 \times 10^7$ & No & --\\
IR35 & 25 & 45 & 10 & log & $2.0 \times 10^7$ & No & --\\
IR65 & 45 & 85 & 10 & log & $2.0 \times 10^7$ & No & --\\
IR100 & 85 & 120 & 10 & log & $2.0 \times 10^7$ & No & --\\
IR160 & 120 & 200 & 10 & log & $2.0 \times 10^7$ & No & --\\
IR300 & 200 & 400 & 10 & log & $2.0 \times 10^7$ & No & --\\
IR600 & 400 & 800 & 10 & log & $2.0 \times 10^7$ & No & --\\
MM & 800 & 1200 & 10 & log & $2.0 \times 10^7$ & No & LABOCA$^{\mathrm{5}}$, JCMT1/2$^{\mathrm{6}}$, IRAM$^{\mathrm{7}}$\\
CM07 & 6000 & 7500 & 10 & log & $2.0 \times 10^7$ & No & VLA$^{\mathrm{8}}$/Q\\
CM13 & 11300 & 16700 & 10 & log & $2.0 \times 10^7$ & No & VLA/K\\
CM20 & 16700 & 25000 & 10 & log & $2.0 \times 10^7$ & No & VLA/Ku\\
CM30 & 25000 & 37500 & 10 & log & $2.0 \times 10^7$ & No & VLA/X\\
CM60 & 37500 & 75000 & 10 & log & $2.0 \times 10^7$ & No & VLA/C\\
\hline\hline
Observable & \multicolumn{2}{c}{$\lambda_{\mathrm{c}}\,[\text{\normalfont\AA}]$} & R & Spacing & $N_{\mathrm{phot}}$ & Image & Comments\\ \hline
H$\alpha$ & \multicolumn{2}{c}{6564.61} & 20,000 & linear & $1.1 \times 10^9$ & Yes & CHARA/SPICA$^{\mathrm{9}}$, R$\sim$10,000\\
H$\beta$ & \multicolumn{2}{c}{4862.71} & 10,000 & linear & $3.0 \times 10^8$ & No & --\\
H$\gamma$ & \multicolumn{2}{c}{4341.69} & 6,000 & linear & $6.5 \times 10^7$ & No & --\\
H$\delta$ & \multicolumn{2}{c}{4102.89} & 6,000 & linear & $6.5 \times 10^7$ & No & --\\
Br11 & \multicolumn{2}{c}{16811.1} & 6,000 & linear & $1.0 \times 10^8$ & No & APOGEE$^{\mathrm{10}}$\\
Br13 & \multicolumn{2}{c}{16113.7} & 6,000 & linear & $1.0 \times 10^8$ & No & APOGEE\\
Br$\gamma$ & \multicolumn{2}{c}{21661.178} & 20,000 & linear & $2.0 \times 10^8$ & Yes & AMBER$^{\mathrm{11}}$, GRAVITY$^{\mathrm{3}}$\\
Pf$\gamma$ & \multicolumn{2}{c}{37405.6} & 10,000 & linear & $1.0 \times 10^8$ & No & --\\
Hu14 & \multicolumn{2}{c}{40208.7} & 10,000 & linear & $1.0 \times 10^8$ & No & --\\
Br$\alpha$ & \multicolumn{2}{c}{40522.6} & 10,000 & linear & $1.0 \times 10^8$ & No & --\\
\hline \hline
\end{tabular}
\begin{tablenotes}
\footnotesize
\item[1] \url{www.eso.org/sci/facilities/paranal/instruments/pionier.html}
\item[2] \url{www.chara.gsu.edu/instrumentation/mirc}
\item[3] \url{www.eso.org/sci/facilities/paranal/decommissioned/midi.html}
\item[4] \url{www.eso.org/sci/facilities/develop/instruments/matisse.html}
\item[5] \url{www.eso.org/public/teles-instr/apex/laboca/}
\item[6] \url{www.eaobservatory.org/jcmt/}
\item[7] \url{www.iram-institute.org/}
\item[8] \url{www.public.nrao.edu/telescopes/vla/}
\item[9] \url{www.chara.gsu.edu/instrumentation/spica}
\item[10] \url{www.sdss4.org/dr12/irspec/}
\item[11] \url{www.eso.org/sci/facilities/paranal/instruments/amber/overview.html}
\end{tablenotes}
\end{threeparttable}
\end{table}

\newpage

\begin{figure}
    \centering
    \includegraphics[scale=0.95]{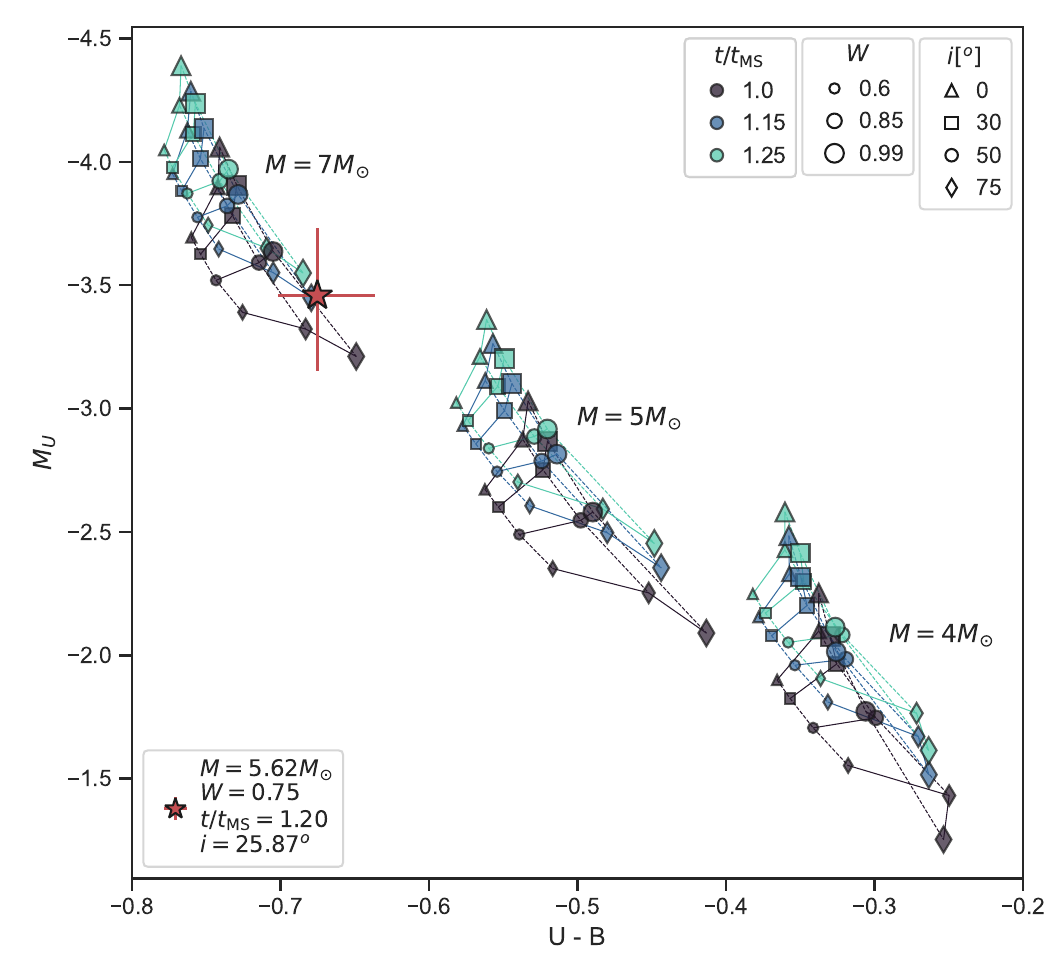}
    \caption{Colour-magnitude diagram for selected models from the Geneva grid for three different masses, from $4$M$_\odot$ to $7$M$_\odot$. Different markers indicate inclinations, marker size indicates rotation rate W and marker colour indicates \ttms. The red star marks the most likely solution given by the UV-only simulation for $\alpha$ Eri, as per Tab. \ref{tab:aeri_res}.}
\label{fig:full_genebra}
\end{figure}

\clearpage
\section{\bcmi polarimetric data and H$\alpha$ profiles}\label{app:obs_data_pol}
\clearpage
\begin{table*}
\begin{threeparttable}
\centering
\small
\caption{Polarimetric data from OPD-LNA taken with the IAGPOL polarimeter.}
\label{tab:polarimetric_bcmi_data}
\begin{tabular}{l l l r l r}
\hline
\hline
MJD         & Filter    & $P_{\mathrm{obs}} [\%]$   & $PA_{\mathrm{Pobs}}$  & $\sigma_P [\%]$ 		& $\sigma_{PA}$  \\ \hline
55855.33	&	B		&	0.0365					&	65.33				&	0.0038				&	2.98   \\
55855.34	&	R		&	0.0512					&	88.90				&	0.0040				&	2.24   \\
55855.37	&	I		&	0.0714					&	78.06				&	0.0195				&	7.82   \\
56022.03	&	I		&	0.0488					&	1.53				&	0.0151				&	8.87   \\
56022.04	&	B		&	0.0194					&	72.54				&	0.0129				&	19.05  \\
56022.05	&	V		&	0.0288					&	92.40				&	0.0152				&	15.12  \\
56022.06	&	R		&	0.0425					&	17.03				&	0.0171				&	11.53  \\
56610.19	&	V		&	0.0488					&	3.48				&	0.0111				&	6.52   \\
56610.22	&	I		&	0.0567					&	0.00				&	0.0042				&	2.12   \\
56610.23	&	R		&	0.0526					&	3.66				&	0.0242				&	13.18  \\
56610.25	&	B		&	0.0637					&	1.90				&	0.0351				&	15.79  \\
56714.09	&	V		&	0.1242					&	85.81				&	0.0127				&	2.93   \\
56726.08	&	B		&	0.0601					&	75.52				&	0.0260				&	12.39  \\
56726.09	&	V		&	0.0410					&	75.25				&	0.0233				&	16.28  \\
56726.10	&	R		&	0.0656					&	80.67				&	0.0239				&	10.44  \\
56726.11	&	I		&	0.0379					&	5.24				&	0.0209				&	15.8   \\
56946.28	&	B		&	0.0826					&	86.66				&	0.0070				&	2.43   \\
56946.30	&	V		&	0.0401					&	106.55				&	0.0152				&	10.86  \\
56946.31	&	R		&	0.0601					&	79.15				&	0.0084				&	4.00   \\
56946.32	&	I		&	0.0306					&	1.22				&	0.0136				&	12.73  \\
56981.29	&	V		&	0.0823					&	3.00				&	0.0267				&	9.29   \\
56981.30	&	R		&	0.0624					&	6.25				&	0.0089				&	4.09   \\
56981.30	&	I		&	0.0505					&	12.85				&	0.0144				&	8.17   \\ \hline
\end{tabular}
\end{threeparttable}
\end{table*}

\begin{figure*}
    \centering
    \includegraphics[scale=.4]{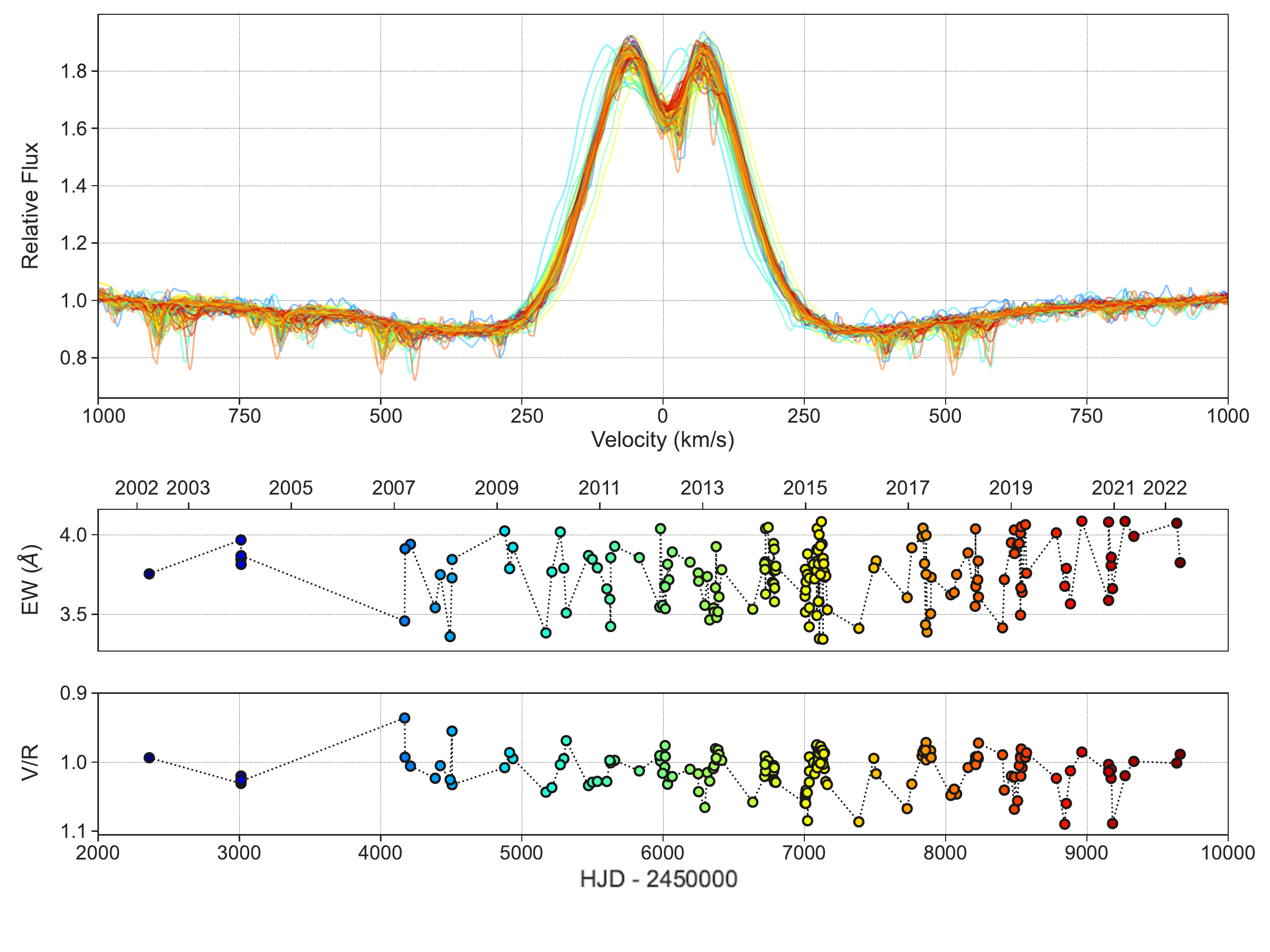}
    \caption{{Top panel: H$\alpha$ profiles from the BeSS database, coloured according to the dates shown in the other panels. Middle panel: Equivalent width (EW) measurements of all profiles. In the top axis, dates are shown in years. Bottom panel: Violet over red ratios (V/R) calculated for all profiles. In the bottom axis, dates are shown in shifted Heliocentric Julian Dates. Both EW and V/R panels share the same time scale.}}
    \label{fig:bcmi_bess}
\end{figure*}

\clearpage
\newpage

\section{Additional corner plots}\label{sect:add_corners}
\clearpage

\begin{figure*}
    \centering
    \includegraphics[scale=0.43]{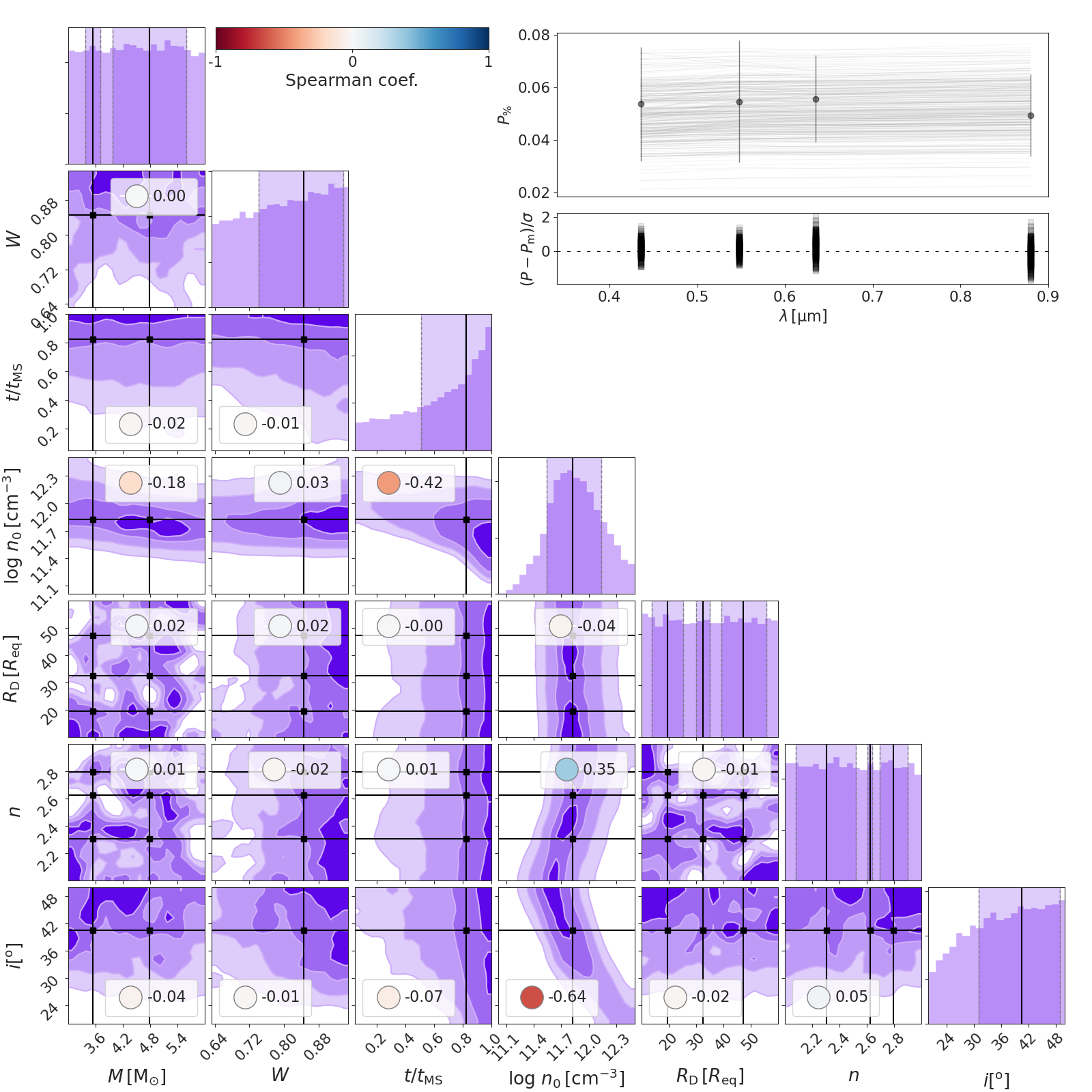}
    \caption{Same as Fig. \ref{fig:bcmi_full}, but for the polarisation data of \bcmi.}
    \label{fig:bcmi_pol}
\end{figure*}

\newpage
\begin{figure*}
    \centering
    \includegraphics[scale=0.33]{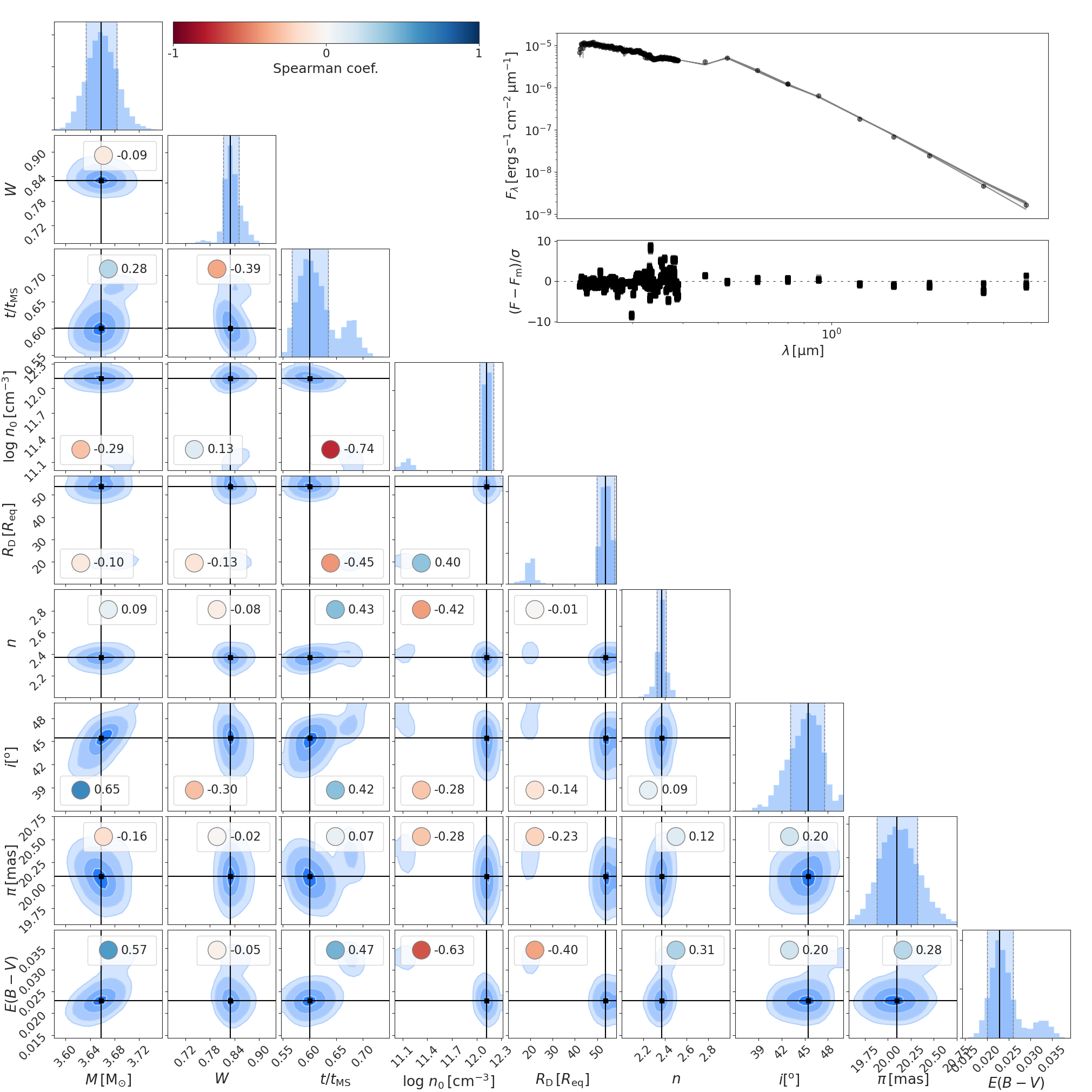}
    \caption{Same as Fig. \ref{fig:bcmi_full}, but for the UV-NIR section of the SED data of \bcmi.}
    \label{fig:bcmi_uv-nir}
\end{figure*}

\newpage
\begin{figure*}
    \centering
    \includegraphics[scale=0.33]{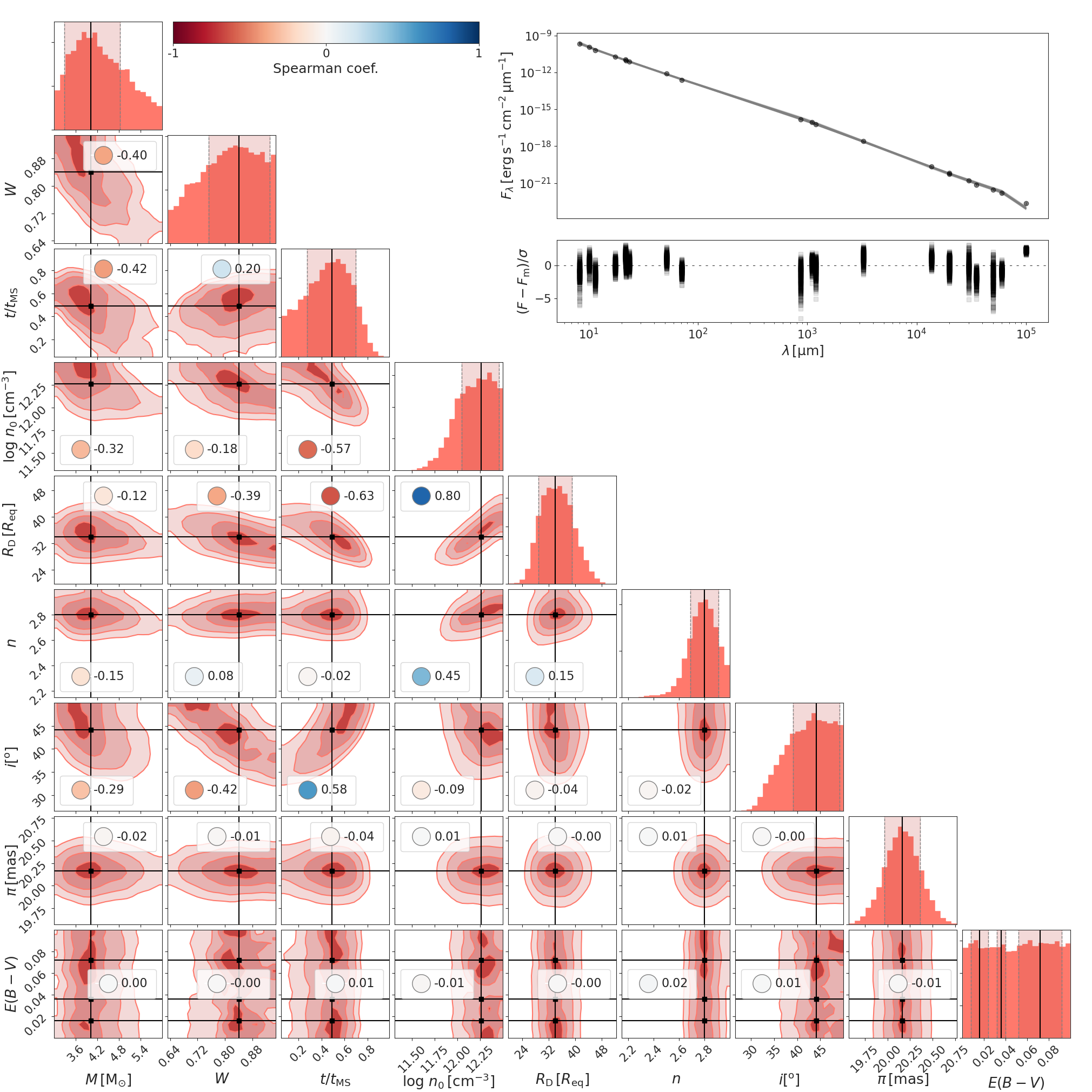}
    \caption{Same as Fig. \ref{fig:bcmi_full}, but for the MIR-Radio section of the SED data of \bcmi.}
    \label{fig:bcmi_mir-radio}
\end{figure*}


\clearpage

\section{Photospheric grid on a Colour-Magnitude diagram}\label{sect:observationalHR}

\clearpage
\begin{figure*}[H]
\centering
\includegraphics[width=1\linewidth]{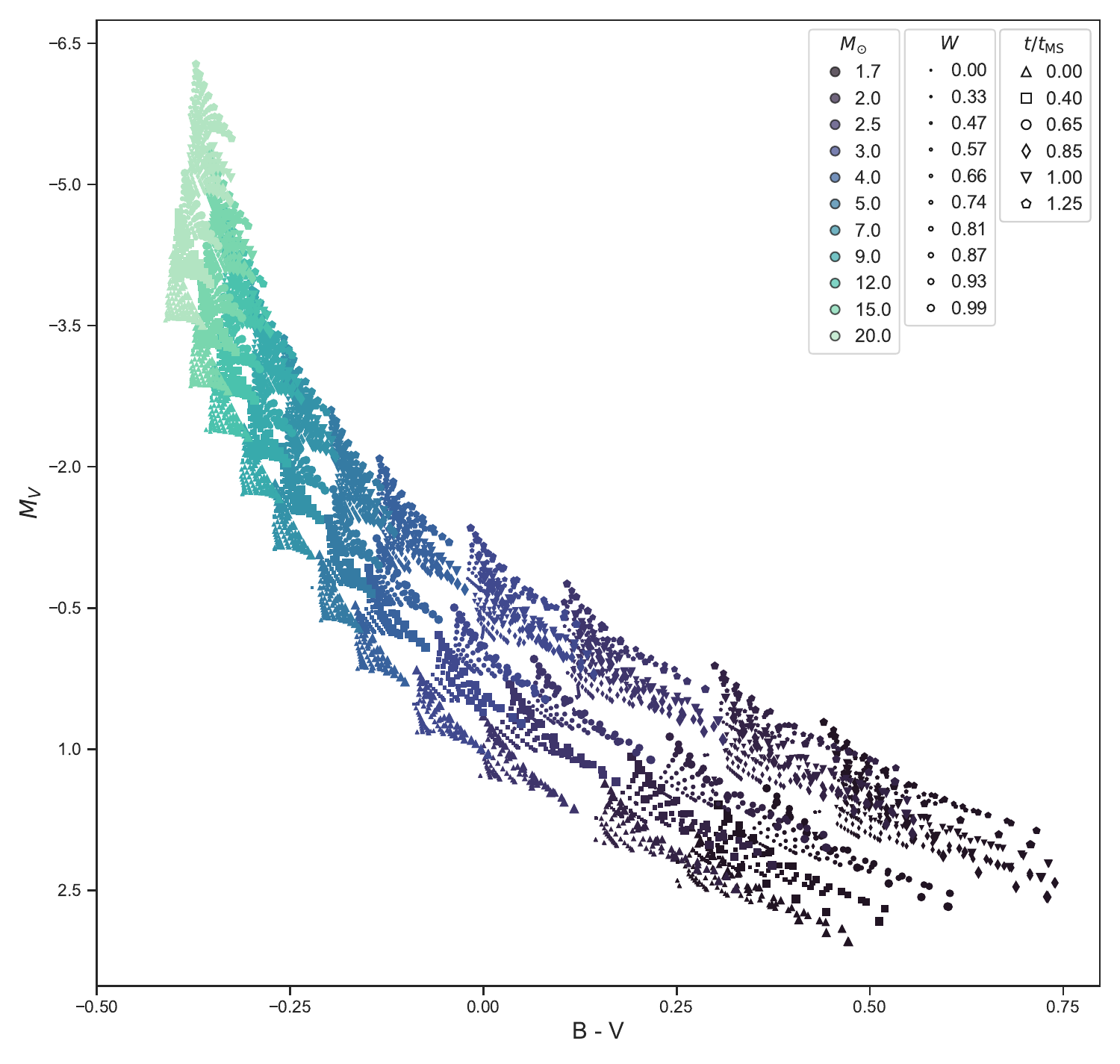}
\caption{
Absolute magnitude $\mathrm{M}_\mathrm{V}$ $\times$ (B-V) colour-magnitude diagram of the photospheric grid on the Vega system. The configuration of this figure (markers' colours, types and sizes) follows Fig. \ref{fig:HRphot}. A significant feature shows up on the observational diagram, which is the cone opening caused by the different inclination angles. Models observed pole-on, when rotating fast, seem brighter, while models observed edge-on seem dimmer and redder.}
\label{fig:HRobservational}
\end{figure*}

\end{appendix}
\bsp	
\label{lastpage}
\end{document}